\documentclass[12pt]{article}
\usepackage{fullpage,amsmath,amsfonts,graphicx,amsthm,url}
\usepackage[matrix,arrow]{xy}
\usepackage{lscape}
\usepackage{subfigure}
\usepackage{multicol}
\usepackage{multirow}
\usepackage{epstopdf}
\usepackage{bm}
\usepackage{natbib}

\theoremstyle{definition}

\def\H{{\bf{H}}}
\def\h{{\bf{h}}}
\def\x{{\bf{x}}}
\def\y{{\bf{y}}}
\def\z{{\bf{z}}}

\def\w{{\bf{w}}}
\def\T{{\bf{T}}}

\def\Z{{\bf{Z}}}
\def\P{{\bf{P}}}
\def\e{{\bf{e}}}
\def\T{{\bf{T}}}
\def\S{{\bf{S}}}
\def\C{{\bf{C}}}

\def\A{{\bf{A}}}
\def\V{{\bf{V}}}
\def\D{{\bf{D}}}

\def\Gam{{\bf{\Gamma}}}

\def\L{{\bm{L}}}

\def\xobs{{\bf{\y_{\rm{o}}}}}
\def\zobs{{\bf{\y_{\rm{o}}}}}
\def\Nobs{N_{\rm{obs}}}

\def\za{\z_a}
\def\zf{\z_f}

\def\Id{{\bf{I}}}
\def\tobs{t_{\rm{obs}}}
\def\robs{{\bf{r}}_{\rm{o}}}
\def\Robs{{\bf{R}}_{\rm{o}}}
\def\RobsI{{\bf{R}}^{-1}_{\rm{o}}}

\def\Pf{{\bf{P}}_{\rm{f}}}
\def\PfI{{\bf{P}}^{-1}_{\rm{f}}}

\def\Rw{{\bf{R}}_{\rm{w}}}
\def\RwI{{\bf{R}}^{-1}_{\rm{w}}}
\def\atarget{{\bf{A}}_{\rm{clim}}}
\def\atargetI{{\bf{A}}^{-1}_{\rm{clim}}}
\def\amtar{{\bf{a}}_{\rm{clim}}}
\def\Kr{{\bf{K}}_{\rm{w}}}
\def\KR{{\bf{K}}_{\rm{o}}}
\def\Bstar{{\bm{{\mathcal{P}}}}}
\def\BstarI{{\bm{{\mathcal{P}}}}^{-1}}
\def\Pf{{\bf{P}}_f}
\def\PfI{{\bf{P}}_f^{-1}}
\def\sigmaclim{\sigma_{\rm{clim}}}
\def\muclim{\mu_{\rm{clim}}}

\begin{document}

\title{Controlling overestimation of error covariance in ensemble Kalman filters with sparse observations: A variance limiting Kalman filter}

\author{\textsc{Georg A. Gottwald}
	\thanks{\textit{Corresponding author:} georg.gottwald@sydney.edu.au}
	\quad\textsc{and Lewis Mitchell}\\ 
	\textit{\footnotesize{School of Mathematics and Statistics,
			             University of Sydney, 
                                          NSW 2006, Australia. }}
\and 
\centerline{\textsc{Sebastian Reich}}\\% Add additional authors, different insitution
\centerline{\textit{\footnotesize{Universit\"at Potsdam, 
Institut f\"ur Mathematik, Am Neuen Palais 10,
D-14469 Potsdam, Germany. 
}}}
}

\maketitle

%%%%%%%%%%%%%%%%%%%%%%%%%%%%%%%%%%%%%%%%%%%

\begin{abstract}
We consider the problem of an ensemble Kalman filter when only partial observations are available. In particular we consider the situation where the observational space consists of variables which are directly observable with known observational error, and of variables of which only their climatic variance and mean are given. To limit the variance of the latter poorly resolved variables we derive a variance limiting Kalman filter (VLKF) in a variational setting. We analyze the variance limiting Kalman filter for a simple linear toy model and determine its range of optimal performance. We explore the variance limiting Kalman filter in an ensemble transform setting for the Lorenz-96 system, and show that incorporating the information of the variance of some un-observable variables can improve the skill and also increase the stability of the data assimilation procedure. 
\end{abstract}

%%%%%%%%%%%%%%%%%%%%%%%%%%%%%%%%%%%%%%%%%%%

\section{Introduction}
In data assimilation one seeks to find the best estimation of the state of a dynamical system given a forecast model with a possible model error and noisy observations at discrete observation intervals \citep{Kalnay}. This process is complicated on the one hand by the often chaotic nature of the underlying nonlinear dynamics leading to an increase of the variance of the forecast, and on the other hand by the fact that one often has only partial information of the observables. In this paper we address the latter issue. We consider situations whereby noisy observations are available for some variables but not for other unresolved variables. However, for the latter we assume that some prior knowledge about their statistical climatic behaviour such as their variance and their mean is available.\\ 

A particularly attractive framework for data assimilation are ensemble Kalman filters (see for example \cite{Evensen}). These straightforwardly implemented filters distinguish themselves from other Kalman filters in that the spatially and temporally varying background error covariance is estimated from an ensemble of nonlinear forecasts. Despite the ease of implementation and the flow-dependent estimation of the error covariance ensemble Kalman filters are subject to several errors and specific difficulties (see \cite{Ehrendorfer07} for a recent review). Besides the problems of estimating model error which is inherent to all filters, and inconsistencies between the filter assumptions and reality such as non-Gaussianity which render all Kalman filters suboptimal, ensemble based Kalman filters have the specific problem of sampling errors due to an insufficient size of the ensemble. These errors usually underestimate the error covariances which may ultimately lead to filter divergence when the filter trusts its own forecast and ignores the information given by the observations.\\

To counteract the associated small spread of the ensemble several techniques have been developed. To deal with errors in ensemble filters due to sampling errors we mention two of the main algorithms, covariance inflation and localisation. To avoid filter divergence due to an underestimation of error covariances the concept of covariance inflation was introduced whereby the prior forecast error covariance is increased by an inflation factor \citep{AndersonAnderson99}. This is usually done in a global fashion and involves careful and expensive tuning of the inflation factor; however recently methods have been devised to adaptively estimate the inflation factor from the innovation statistics \citep{Anderson07,Anderson09,Li09}. Too small ensemble sizes also lead to spurious correlations associated with remote observations. To address this issue the concept of localization has been introduced \citep{Houtekamer98,Houtekamer01,Hamill01,Ott04,Szunyogh05} whereby only spatially close observations are used for the innovations.\\

To take into account the uncertainty in the model representation we mention here isotropic model error parametrization \citep{MitchellHoutekamer00,HoutekamerEtAl05}, stochastic parametrizations \citep{Buizza99} and kinetic energy backscatter \citep{Shutts05}. A recent comparison between those methods is given in \cite{Houtekamer09,Charron10}, see also \cite{HamillWhitaker05}. The problem of non-Gaussianity is for example discussed in \cite{Pires10,Bocquet10}.\\

Whereas the underestimation of error covariances has received much attention, relatively little is done for a possible overestimation of error covariances. Overestimation of covariance is a finite ensemble size effect which typically occurs in sparse observation networks (see for example \cite{Liuetal08,Whitakeretal09}). Uncontrolled growth of error covariances which is not tempered by available observations may progressively spoil the overall analysis. This effect is even exacerbated when inflation is used; in regions where no observations influence the analysis, inflation can lead to unrealistically large ensemble variances progressively degrading the overall analysis (see for example \cite{Whitakeretal04}). This is particularly problematic when inappropriate uniform inflation is used. Moreover, it is well known that covariance localization can be a significant source of imblance in the analyzed fields (see for example \cite{HoutekamerMitchell05,Kepert09,Houtekamer09}). Localization artificially generates unwanted gravity wave activity which in poorly resolved spatial regions may lead to an unrealistic overestimation of error covariances. Being able to control this should help filter performances considerably.\\

When assimilating current weather data in numerical schemes for the troposphere, the main problem is underestimation of error covariances rather than overestimation. This is due to the availability of radiosonde data which assures wide observational coverage. However, in the pre-radiosonde era there were severe data voids, particularly in the southern hemisphere and in vertical resolution since most observations were done on the surface level in the northern hemisphere. There is an increased interest in so called climate reanalysis (see for example \citep{Bengtson07,Whitakeretal04}), which has the challenge to deal with large unobserved regions. Historical atmospheric observations are reanalyzed by a fixed forecast scheme to provide a global homogeneous dataset covering troposphere and stratosphere for very long periods. A remarkable effort is the international \emph{Twentieth Century Reanalysis Project} (20CR) \citep{Compoetal11}, which produced a global estimate of the atmosphere for the entire 20th century (1871 to the present) using only synoptic surface pressure reports and monthly sea-surface temperature and sea-ice distributions. Such a dataset could help to analyze climate variations in the twentieth century or the multidecadal variations in the behaviour of the El-Ni\~no-Southern Oscillation. An obstacle for reanalysis is the overestimation of error covariances if one chooses to employ ensemble filters (\cite{Whitakeretal04} where multiplicative covariance inflation is employed).\\
%A large obstacle for reanalysis are overestimations of error covariances if one chooses to employ ensemble filter methods \citep{Whitakeretal04}.\\

Overestimation of error covariances occurs also in modern numerical weather forecast schemes for which the upper lid of the vertical domain is constantly pushed towards higher and higher levels to incorporate the mesosphere, with the aim to better resolve processes in the polar stratosphere (see for example \cite{Polavarapu05,Sankey07,Eckermann09}). The energy spectrum in the mesosphere is, contrary to the troposphere, dominated by gravity waves. The high variability associated with these waves causes very large error covariances in the mesosphere which can be $2$ orders of magnitude larger than at lower levels \citep{Polavarapu05}, rendering the filter very sensitive to small uncertainties in the forecast covariances. Being able to control the variances of mesospheric gravity waves is therefore a big challenge.\\

The question we address in this work is how can the statistical information available for some data which are otherwise not observable, be effectively incorporated in data assimilation to control the potentially high error covariances associated with the data void. We will develop a framework to modify the familiar Kalman filter (see for example \citep{Evensen,Simon}) for partial observations with only limited information on the mean and variance, with the effect that the error covariance of the unresolved variables cannot exceed their climatic variance and their mean is controlled by driving it towards the climatological value.\\

The paper is organized as follows. In Section~\ref{sec-formulation} we will introduce the dynamical setting and briefly describe the ensemble transform Kalman filter (ETKF), a special form of an ensemble  square root filter. In Section~\ref{sec-deriv} we will derive the variance limiting Kalman filter (VLKF) in a variational setting. In Section~\ref{sec-toymodel} we illustrate the VLKF with a simple linear toy model for which the filter can be analyzed analytically. We will extract the parameter regimes where we expect VLKF to yield optimal performance. In Section~\ref{sec-numerics} we apply the VLKF to the $40$-dimensional Lorenz-96 system \citep{Lorenz96} and present numerical results illustrating the advantage of such a variance limiting filter. We conclude the paper with a discussion in section~\ref{sec-disc}.

%%%%%%%%%%%%%%%%%%%%%%%%%%%%%%%%%%%%%%%%%%%

\section{Setting}
\label{sec-formulation}
Assume an $N$-dimensional\footnote{The exposition is restricted to $\mathbb{R}^N$, but we note that the formulation can be generalized for Hilbert spaces.} dynamical system whose dynamics is given by
\begin{equation}
\label{e.ode}
\dot{\bf z} = f({\bf z})\;,
\end{equation}
with the state variable ${\bf z}\in \mathbb{R}^N$. We assume that the state space is decomposable according to ${\bf z}=({\bf x},{\bf y})$ with ${\bf x}\in \mathbb{R}^n$ and ${\bf y}\in \mathbb{R}^m$ and $n+m=N$. Here ${\bf x}$ shall denote those variables for which direct observations are available, and ${\bf y}$ shall denote those variables for which only some integrated or statistical information is available. We will coin the former \emph{observables} and the latter \emph{pseudo-observables}. We do not incorporate model error here and assume that (\ref{e.ode}) describes the truth. We apply the notation of \cite{Ide97} unless stated explicitly otherwise.

Let us introduce an observation operator $\H:\mathbb{R}^N\to \mathbb{R}^n$ which maps from the whole space into observation space spanned by the designated variables $\x$. We assume that observations of the designated variables $\x$ are given at equally spaced discrete observation times $t_i$ with the observation interval $\Delta\tobs$. Since it is assumed that there is no model error, the observations ${\xobs}\in \mathbb{R}^n$ at discrete times $t_i=i
\Delta \tobs$ are given by
\[
{\xobs}(t_i) = \H \z(t_i) + {\robs}\, ,
\]
with independent and identically distributed ({\rm i.i.d.}) observational Gaussian noise $\robs \in \mathbb{R}^n$. The observational noise is assumed to be independent of the system state, and to have zero mean and constant covariance $\Robs\in \mathbb{R}^{n\times n}$.

We further introduce an operator $\h:\mathbb{R}^N \to \mathbb{R}^m$ which maps from the whole space into the space of the pseudo-observables spanned by $\y$. We assume that the pseudo-observables have variance $\atarget \in \mathbb{R}^{m\times m}$ and constant mean $\amtar \in \mathbb{R}^{m}$. This is the only information available for the pseudo-observables, and may be estimated, for example, from climatic measurements. The error covariance of those pseudo-observations is denoted by $\Rw \in \mathbb{R}^{m\times m}$.

The model forecast state $\zf$ at each observation interval is obtained by integrating the state variable with the full nonlinear dynamics (\ref{e.ode}) for the time interval $\Delta \tobs$. The background (or forecast) involves an error with covariance $\Pf\in \mathbb{R}^{N\times N}$.

Data assimilation aims to find the best estimation of the current state given the forecast $\zf$ with variance $\Pf$ and observations $\xobs$ of the designated variables with error covariance $\Robs$. Pseudo-observations can be included following the standard Bayesian approach once their mean $\amtar$ and error covariance $\Rw$ are known. However, the error covariance $\Rw$ of a pseudo-observation is in general not equal to $\atarget$. In Section~\ref{sec-deriv}, we will show how to derive the error  covariance $\Rw$ in order to ensure that the forecast does not exceed the prescribed variance $\atarget$. We do so in the framework of  Kalman filters and shall now briefly summarize the basic ideas to construct such a filter for the case of an ensemble square root filter \citep{Tippett03}, the ensemble transform filter \citep{Wang04}.

%%%%%%%%%%%%%%%%%%%%%%%%%%%%%%%%%%%%%%%%%%

\subsection{Ensemble Kalman filter}
\label{sec-EnKF}
In an ensemble Kalman filter (EnKF) \citep{Evensen} an ensemble with $k$ members $\z_k$
\[
\Z=\left[ \z_1,\z_2,\dots,\z_k \right]
\in \mathbb{R}^{N\times k}
\]
is propagated by the full nonlinear dynamics (\ref{e.ode}), which is written as
\begin{equation}
{\dot{\Z}} = {\bf{f}}(\Z)\; ,
\qquad 
{\bf{f}}(\Z) =\left[ f(\z_1),f(\z_2),\dots,f(\z_k) \right]
\in \mathbb{R}^{N\times k} \; .
\label{e.odeE}
\end{equation}
The ensemble is split into its mean 
\[
{\bar{\z}} = \frac{1}{k}\sum_{i=1}^k\z_i=\Z\w
\qquad {\rm{with}} \qquad
\w =
\frac{1}{k}\e
\in  \mathbb{R}^{k}
\; ,
\]
where $\e=\left[1,\dots,1\right]^T \in \mathbb{R}^{k}$, and its ensemble deviation matrix
\[
\Z^\prime=\Z-{\bar{\z}}\e^T=\Z\T \; ,
\]
with the constant projection matrix 
\[
\T = \Id-\w\e^T 
\in \mathbb{R}^{k\times k}
\; .
\]
The ensemble deviation matrix $\Z^\prime$ can be used to approximate the ensemble forecast covariance matrix via
\[
\P_f(t)
= 
\frac{1}{k-1}\Z^\prime(t)\left[\Z^\prime(t)\right]^T
\in \mathbb{R}^{N\times N}\; .
\]
%Note that $\P_f(t)$ is rank-deficient for $k<N$ which is the typical situation in numerical weather prediction where $N$ is of the order of $10^9$ and $k$ of the order of $100$.
%Note that $\P_f(t)$ is rank-deficient if the ensemble size $k$ is smaller than the rank of the covariance matrix. The rank is generically not known in atmospheric models with $N$ of the order of $10^9$, but is believed to be orders of magnitudes smaller than $N$ and orders of magnitude larger than $100$, the typical ensemble size in numerical weather prediction.

Given the forecast ensemble $\Z_f=\Z(t_i-\epsilon)$ and the associated forecast error covariance matrix (or the \emph{prior}) $\P_f(t_i-\epsilon)$, the actual Kalman analysis \citep{Kalnay,Evensen,Simon} updates a forecast into a so-called analysis (or the \emph{posterior}). Variables at times $t=t_i - \epsilon$ are evaluated before taking the observations (and/or pseudo-observations) into account in the analysis step, and variables at times $t=t_i + \epsilon$ are evaluated after the analysis step when the observations (and/or pseudo-observations) have been taken into account. In the first step of the analysis the forecast mean,
\[
{\bar{\z}}_f = \Z_f\w\;,
\]
is updated to the analysis mean
\begin{eqnarray}
\label{e.zaens}
{\bar{\z}}_a 
= 
{\bar{\z}}_f 
- \KR\left[\H{\bar{\z}}_f - {\zobs} \right] 
- \Kr \left[\h{\bar{\z}}_f - \amtar \right]\;,
\end{eqnarray}
where the Kalman gain matrices are defined as
\begin{eqnarray}
\KR &=& \P_a\H^T\RobsI
\nonumber \\
\Kr &=& \P_a\h^T\RwI\; .
\label{e.KGM}
\end{eqnarray}
The analysis covariance $\P_a$ is given by the addition rule for variances, typical in linear Kalman filtering
\citep{Kalnay},
\begin{eqnarray}
\label{e.Pa}
\P_a = \left(\PfI + \H^T\RobsI \H + \h^T\RwI \h \right)^{-1} 
\, .
\end{eqnarray}
To calculate an ensemble $\Z_a$ which is consistent with the error covariance after the observation $\P_a$, and which therefore needs to satisfy
\[
\P_a
= 
\frac{1}{k-1}\Z_a\T\left[\Z_a\right]^T\; ,
\]
we use the method of ensemble square root filters \citep{Simon}. In particular we use the method proposed in \citep{Tippett03,Wang04}, the so called ensemble transform Kalman filter (ETKF), which seeks a transformation $\S \in \mathbb{R}^{k\times k}$ such that
\begin{equation}
\Z_a^\prime=\Z_f^\prime \S \; .
\label{e.S}
\end{equation}
Alternatively one could have chosen the ensemble adjustment filter \citep{Anderson01} in which the ensemble deviation matrix $\Z_f^\prime$ is pre-multiplied with an appropriately determined matrix $\A\in \mathbb{R}^{N\times N}$. However, since we are mainly interested in the case $k\ll N$ we shall use the ETKF. Note that the matrix $\S$ is not uniquely determined for $k<N$. The transformation matrix $\S$ can be obtained either by using continuous Kalman filters \citep{BGR09} or directly \citep{Wang04} by 
\[
\S = 
{\bar{\C}}
\left(\Id_k + {\bar{\Gam}} \right)^{-\frac{1}{2}} 
{\bar{\C}}^T\;.
\]
Here $\C\Gam\C^T$ is the singular value decomposition of
\[
{\bf{U}} 
= \frac{1}{k-1}
\T^T\Z_f^T
\left( \H^T\RobsI\H + \h^T\RwI\h\right)
\Z_f\T\; .
\]
The matrix ${\bar{\C}}\in \mathbb{R}^{k\times (k-1)}$ is obtained by erasing the last zero column from $\C\in\mathbb{R}^{k\times k}$, and ${\bar{\Gam}} \in\mathbb{R}^{(k-1)\times (k-1)}$ is the upper left $(k-1)\times(k-1)$ block of the diagonal matrix $\Gam\in\mathbb{R}^{k\times k}$. The deletion of the $0$ eigenvalue and the associated columns in $\C$ assure that $\Z_a^\prime = \Z_a^\prime \S$ and therefore that the analysis mean is given by ${\bar{\z}}_a$. Note that $\S$ is symmetric and $\S \T=\T \S$ which assures that $\Z_a^\prime = \Z_a^\prime \S$ implying that the mean is preserved under the transformation. This is not necessarily true for general ensemble transform methods of the form (\ref{e.S}).

A new forecast $\Z(t_{i+1} -\epsilon)$ is then obtained by propagating $\Z_a$ with the full nonlinear dynamics (\ref{e.odeE}) to the next time of observation. The numerical results presented later in Sections~\ref{sec-toymodel} and \ref{sec-numerics} are obtained with this method. 

In the next Section we will determine how the error covariance $\Rw$ used in the Kalman filter is linked to the variance $\atarget$ of the pseudo-variables.

%%%%%%%%%%%%%%%%%%%%%%%%%%%%%%%%%%%%%%%%%%

\section{Derivation of the variance limiting Kalman filter}
\label{sec-deriv}
One may naively believe that the error covariance of the pseudo-observable $\Rw$ is determined by the target variance of the pseudo-observables  $\atarget$ simply by setting $\Rw=\atarget$. In the following we will see that this is not true, and that the expression for $\Rw$ which ensures that the variance of the pseudo-observables in the analysis is limited from above by $\atarget$ involves all error covariances.

%It is well known that Kalman filtering is equivalent to the variational 3DVAR method which aims at minimizing a cost function \citep{Kalnay}. 
We formulate the Kalman filter as a minimization problem of a cost function (e.g. \cite{Kalnay}). The cost function for one analysis step as described in Section~\ref{sec-EnKF} with a given background $\zf$ and associated error covariance $\Pf$ is typically written as
\begin{eqnarray}
J({\z}) &=& 
 \frac{1}{2}({\z}-{\z}_f)^T\PfI(\z-\z_f)
+ \frac{1}{2}(\zobs-\H\z)^T\RobsI(\zobs-\H\z) 
\nonumber
\\
&&
+\; \frac{1}{2}(\amtar-\h\z)^T\RwI(\amtar-\h\z)
\; ,
\end{eqnarray}
where $\z$ is the state variable at one observation time $t_i=i \Delta \tobs$. Note that the part involving the pseudo-observables corresponds to the notion of {\it weak constraints} in variational data assimilation \citep{Sasaki70,Zupanski97,Neef02}. 

The analysis step of the data assimilation procedure consists of finding the critical point of this cost function. The thereby obtained analysis ${\z={\bar \z}_a}$ and the associated variance $\P_a$ are then subsequently propagated to the next observation time $t_{i+1}$ to yield ${\z}_f$ and $\Pf$ at the next time step, at which a new analysis step can be performed. The equation for the critical point with $\nabla_{{\z}}J({\z})=0$ is readily evaluated to be
\begin{eqnarray}
\left( \PfI + {\H}^T\RobsI \H + {\h}^T\RwI \h \right)\za 
= 
\PfI \zf + {\H}^T\RobsI {\zobs} 
+ \h^T\RwI \amtar
\; ,
\end{eqnarray}
and yields (\ref{e.zaens}) for the analysis mean ${\bar \z}_a$, and (\ref{e.Pa}) for the analysis covariance $\P_a$ with Kalman gain matrices given by (\ref{e.KGM}). 

%which can be written as
%\begin{eqnarray}
%\label{e.za}
%\za
%= 
%\zf - \KR\left[\H{\zf} -{\zobs} \right] - \Kr \left[\h{\zf} - \amtar \right]\;,
%\end{eqnarray}
%where
%\begin{eqnarray}
%\KR &=& \P_a\H^T\RobsI
%\nonumber \\
%\Kr &=& \P_a\h^T\RwI\; ,
%\end{eqnarray}
%with the covariance of the analysis
%\begin{eqnarray}
%\label{e.Pa2}
%\P_a = \left(\PfI + \H^T\RobsI \H + \h^T\RwI \h \right)^{-1} 
%\; .
%\end{eqnarray}
%The last equation can be used directly for ensemble Kalman filters (see (\ref{e.Pa})). The equations (\ref{e.za}) for the state variables $\za$ have to be formulated for the mean variables in the setting of ensemble Kalman filters (compare with (\ref{e.zaens})).

To control the variance of the unresolved pseudo-observables $\amtar=\h \z$ we set
\begin{equation}
\label{e.Paconst}
\h \P_a\h^T = \atarget\;.
\end{equation}
%where $\P_a^+$ denotes the covariance propagated with the dynamics (\ref{e.ode}) for one cycle time $\Delta \tobs$. 
Introducing
\begin{equation}
\label{e.BI}
%\BstarI = \PfIplus + \H^T\RobsI\H\; ,
\BstarI = \PfI + \H^T\RobsI\H\; ,
\end{equation}
and upon applying the Sherman-Morrison-Woodbury formula (see for example \cite{Golub}) to $(\BstarI + \h^T\RwI \h)^{-1}$, equation (\ref{e.Paconst}) yields the desired equation for $\Rw$
\begin{equation}
\label{e.Rw}
\RwI = \atargetI - \left(\h \Bstar \h^T\right)^{-1}  \; , 
\end{equation}
which is yet again a reciprocal addition formula for variances. Note that the naive expectation that $\Rw=\atarget$ is true only for $\Pf \to \infty$, but is not generally true.
% (and also for the trivial case of simultaneous having non-noisy observations with $\Robs \equiv 0$ and no variance of the forecast variance, i.e. $\Pf \equiv 0$). 
For sufficiently small background error covariance $\Pf$, the error covariance $\Rw$ as defined in (\ref{e.Rw}) is not positive semi-definite. In this case the information given by the pseudo-observables has to be discarded. In the language of variational data assimilation the criterion of positive definiteness of $\RwI$ determines whether the weak constraint is switched on or off. To determine those eigendirections for which the statistical information available can be incorporated, we diagonalize $\RwI=\V\D\V^T$ and define ${\bar{\D}}$ with ${\bar{\D}}_{ii}=\D_{ii}$ for $\D_{ii} \ge 0$ and ${\bar{\D}}_{ii}=0$ for $\D_{ii} < 0$. The modified $\RwI=\V{\bar{\D}}\V^T$ then uses information of the pseudo-observables only in those directions which potentially allow for improvement of the analysis. Noting that $\Bstar$ denotes the analysis covariance of an ETKF (with $\Rw=0$), we see that equation (\ref{e.Rw}) states that the variance constraint switches on for those eigendirections whose corresponding singular eigenvalues of $\h \Bstar \h^T$ are larger than those of $\atarget$. Hence the proposed VLKF as defined here incorporates the climatic information of the unresolved variables in order to restrict the posterior error covariance of those pseudo-observables to lie below their climatic variance and to drive the mean towards their climatological mean.

%%%%%%%%%%%%%%%%%%%%%%%%%%%%%%%%%%%%%%%%%%

\section{Analytical linear toy model}
\label{sec-toymodel}
In this Section we study the VLKF for the following coupled linear skew product system for two oscillators $\x\in \mathbb{R}^2$, $\y\in \mathbb{R}^2$
\begin{eqnarray*}
{d \x} &=& {\bm A} \x \, dt - {\bm \Gamma_{\bm x}}\,\x \, dt+ {\bm \sigma_{\bm x}} d{\bm W}_t + {\mathbf \Lambda} \y\, dt\\
{d \y} &=& {\bm B} \y \, dt  - {\bm \Gamma_{\bm y}}\,\y\, dt + {\bm \sigma_{\bm y}} d{\bm B}_t \; ,
\end{eqnarray*}
where $\bm A$, $\bm B$ and $\bm \Lambda$ are all skew-symmetric, $\bm \sigma_{\bm {x,y}}$ and $\bm \Gamma_{\bm {x,y}}$ are all symmetric, and $d{\bm W}_t$ and $d{\bm B}_t$ are independent two-dimensional Brownian processes\footnote{We will use bold font for matrices and vectors, and non-bold font for scalars here. It should be clear from the context whether bold fonts refer to a matrix or a vector.}. We assume here for simplicity that 
\begin{align*}
\bm \Gamma_{\bm x} = \gamma_x \Id\,, \qquad \bm \Gamma_{\bm y} = \gamma_y \Id\,, 
\qquad
\bm \sigma_{\bm x} = \sigma_x \Id \,, \qquad \bm \sigma_{\bm y} = \sigma_y \Id\,, 
\qquad 
\Robs = R_{\rm obs}\Id
\; ,
\end{align*}
with the identity matrix $\Id$, and
\begin{align*}
\bm A = \omega_x \bm J\,,
\qquad
\bm B = \omega_y \bm J\,,
\qquad 
\bm \Lambda = \lambda \bm J
\; ,
\end{align*}
with the skew-symmetric matrix
\begin{align*}
\bm J = 
\left( 
\begin{array}{cc} 
0 & -1\\ 1 & \;\,\,\,0 
\end{array} 
\right) 
\; .
\end{align*}
Note that our particular choice for the matrices implies $\Rw = R_w \Id$.\\

\noindent
The system models two noisy coupled oscillators, $\x$ and $\y$. We assume that we have access to observations of the variable $\x$ at discrete observation times $t_i=i\Delta \tobs$, but have only statistical information about the variable $\y$. We assume knowledge of the climatic mean $\bm \mu_{\rm clim}$ and the climatic covariance $\bm\sigma_{\rm clim}^2$ of the unobserved variable $\y$. The noise is of Ornstein-Uhlenbeck type \citep{Gardiner}, and may represent either model error or parametrize highly chaotic nonlinear dynamics. Without loss of generality, the coupling is chosen such that the $\y$-dynamics drives the $\x$-dynamics but not vice versa. The form of the coupling is not essential for our argument, and it may be oscillatory or damping with $\bm \Lambda = \lambda \Id$. 
%We can associate temperatures for the two equations, $T_x=\sigma_x^2/(2\gamma_x)$ and  $T_y=\sigma_y^2/(2\gamma_y)$. 
We write this system in the more compact form for $\z=(\x,\y)\in \mathbb{R}^4$
\begin{eqnarray}
\label{e.toy}
{d \z} &=& {\bm M} \z \, dt - {\bm \Gamma} \z \, dt+ {\bm \sigma}\, d{\bm W}_t + {\bm{C}}\z\, dt
\end{eqnarray}
with
\begin{align*}
\bm M = 
\left( 
\begin{array}{cc} 
\bm A & \bm 0\\ \bm 0 & \bm B 
\end{array} 
\right) 
\qquad
\bm \Gamma = 
\left( 
\begin{array}{cc} 
\bm \Gamma_{\bm x}& \bm 0\\ \bm 0 & \bm \Gamma_{\bm y}
\end{array} 
\right) 
\\
{\bm \sigma} = 
\left( 
\begin{array}{cc} 
\bm \sigma_x & \bm 0\\ \bm 0 & \bm \sigma_y 
\end{array} 
\right) 
\qquad
{\bm{C}} = 
\left( 
\begin{array}{cc} 
\bm 0 & \bm \Lambda\\ \bm 0 & \bm 0 
\end{array} 
\right) \; .
\end{align*}
The solution of (\ref{e.toy}) can be obtained using It\^o's formula and, introducing the propagator $\L(t) = \exp\left( (\bm M - \bm \Gamma + {\bm C})t \right)$, which commutes with $\bm \sigma$ for our choice of the matrices, is given by
\begin{align*}
\z(t)=\L(t)\z_0 
+ {\bm \sigma} \int_0^t \L(t-s)\, d{\bm W}_s\; ,
\end{align*}
with mean
\begin{align*}
%\bm E \left[ z\right] =e^{(\bm M - \bm \Gamma -{\bm{\mathcal{C}}})t}z_0\; ,
\bm \mu(t) =\L(t)\z_0\; ,
\end{align*}
and covariance
\begin{align}
\label{e.varOU}
{\bm \Sigma}(t)= {\bm \sigma} 
\left(2\bm \Gamma - {\bm{\mathcal{C}}}\right)^{-1}
\left(\Id - \exp\left(-\left(2\bm \Gamma - {\bm{\mathcal{C}}}\right)t\right)\right)
{\bm \sigma}^T\; ,
\end{align}
where
\begin{align*}
{\bm{\mathcal{C}}} = 
\left( 
\begin{array}{cc} 
\bm 0 & \bm \Lambda\\ -\bm \Lambda & \bm 0 
\end{array} 
\right) \; .
\end{align*}
The climatic mean 
%$\mu_{\rm clim}\in \mathbb{R}^4$ 
$\bm \mu_{\rm clim}\in \mathbb{R}^4$
and covariance matrix $\bm \Sigma_{\rm{clim}}\in \mathbb{R}^{4\times 4}$ are then obtained in the limit $t\to\infty$ as
\begin{align*}
%\mu_{\rm clim} = \lim_{t\to\infty}  \bm E \left[ z\right]= 0\; ,
\bm \mu_{\rm clim} = \lim_{t\to\infty}  \bm \mu(t) = 0\; ,
\end{align*}
and 
\begin{align*}
\bm \Sigma_{\rm clim}=\lim_{t\to\infty} {\bm \Sigma}(t) = {\bm \sigma} 
\left(2\bm \Gamma - {\bm{\mathcal{C}}}\right)^{-1}
{\bm \sigma}^T\; .
\end{align*}
In order for the stochastic process (\ref{e.toy}) to have a stationary density and for ${\bm \Sigma}(t)$ to be a positive definite covariance matrix for all $t$, the coupling has to be sufficiently small with $\lambda^2<4 \gamma_x \gamma_y$. Note that the skew product nature of the system (\ref{e.toy}) is not special in the sense that a non-skew product structure where $\x$ couples back to $\y$ would simply lead to a renormalization of ${\bm{\mathcal{C}}}$. However, it is pertinent to mention that although in the actual dynamics of the model (\ref{e.toy}) there is no back-coupling from $\x$ to $\y$, the Kalman filter generically introduces back-coupling of all variables through the inversion of the covariance matrices (cf. (\ref{e.Pa})).\\

\noindent
We will now investigate the variance limiting Kalman filter for this toy model. In particular we will first analyze under what conditions $\Rw$ is positive definite and the variance constraint will be switched on, and second we will analyze when the VLKF yields a skill improvement when compared to the standard ETKF.

\begin{figure}[htbp]
\begin{center}
\includegraphics[width = 0.5\linewidth,height=6cm]{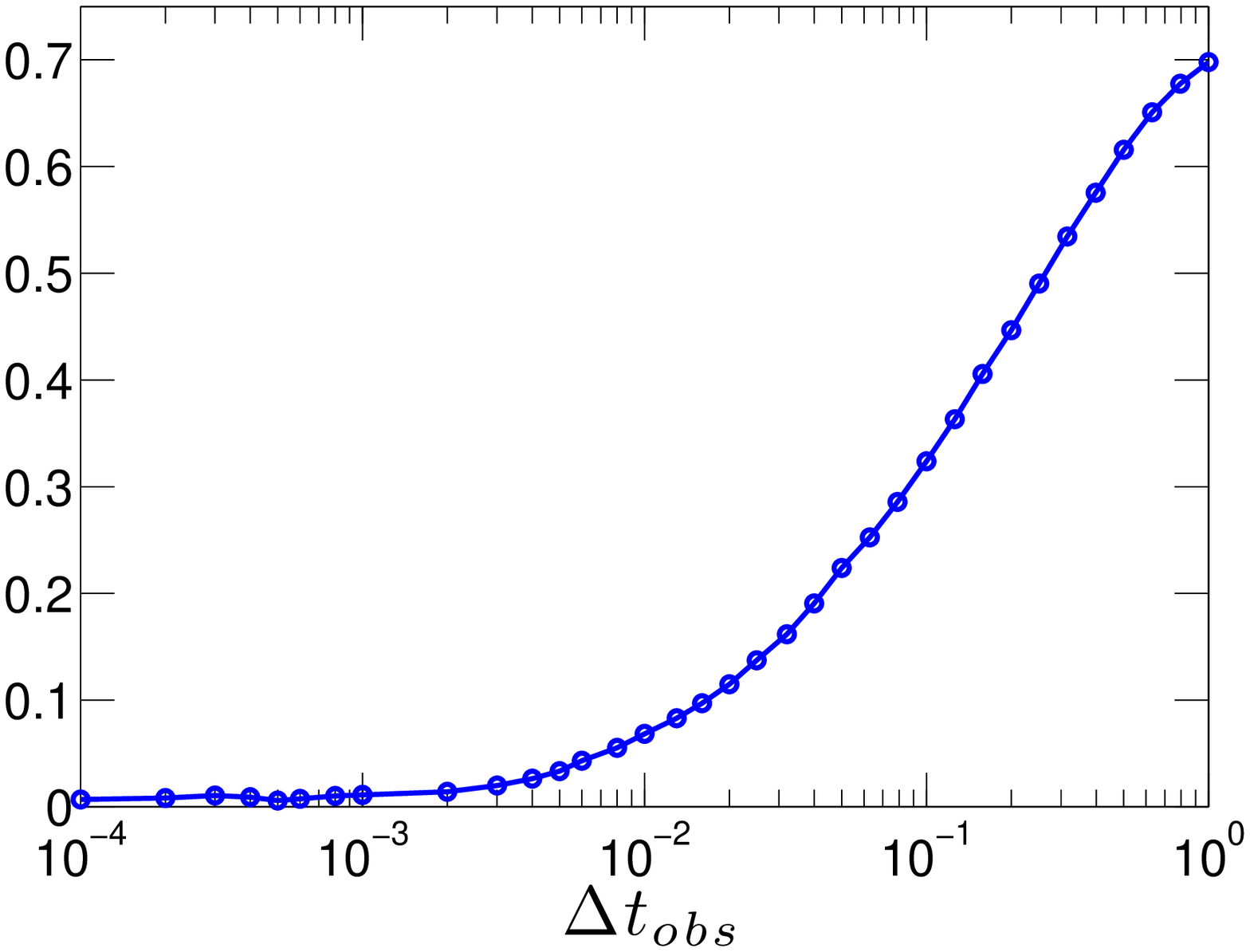}
\caption{Proportion of incidences when the variance constraint is switched on and $\Rw$ is positive definite as a function of the observation interval $\Delta \tobs$ for the stochastic linear toy model (\ref{e.toy}). We used $\gamma_x=1$, $\gamma_y=1$, $\sigma_x=1$, $\sigma_y=1$, $\lambda=0.2$. We used $k=20$ ensemble members, 100 realizations and $\Robs = \H \bm \Sigma_{\rm clim} \H^T$ and no inflation with $\delta=1$. The analytically calculated critical observation interval according to equation (\ref{e.dtobsRw}) is $\Delta \tobs=10^{-2}$.}
\label{fig:dtobstoymodel}
\end{center}
\end{figure}

We start with the positive definiteness of $\Rw$. When calculating the covariance of the forecast in an ensemble filter we need to interpret the solution of the linear toy model (\ref{e.toy}) as
\begin{align*}
{\z}_j(t_{i+1})
%=
\, {\buildrel d \over =}\,
\L(\Delta \tobs) {\z_j}(t_i) 
+ {\bm \sigma} \int_0^{\Delta \tobs} \L (\Delta \tobs-s) d{\bm W_j}_s\qquad j=1,2,\cdots, k\;\, ,
\end{align*}
where $\z_j(t_{i+1})$ is the forecast of ensemble member $j$ at time $t_{i+1}=t_i+\Delta \tobs = (i+1)\Delta \tobs$ before the analysis propagated from its initial condition $\z_j(t_i)={\bar \z}_a(t_i)+\xi_j$ with $\xi_j\sim{\cal{N}}(0,\P_a(t_i))$ at the previous analysis. The equality here is in distribution only, i.e. members of the ensemble are not equal in a pathwise sense as their driving Brownian will be different, but they will have the same mean and variance. The covariance of the forecast can then be obtained by averaging with respect to the ensemble and with respect to realizations of the Brownian motion, and is readily computed as
\begin{align}
\label{e.PfOU}
\P_f(t_{i+1}) = 
\L(\Delta \tobs)\,\P_a(t_i)\, \L^T(\Delta \tobs)
+ {\bm \Sigma}(\Delta \tobs) \; ,
\end{align}
where $\L^T(t)=\exp\left((-\bm M -\bm \Gamma + {\bm C}^T)\,t\right)$ denotes the transpose of $\L(t)$. The forecast covariance of an ensemble with spread $\P_a$ is typically larger than the forecast covariance $\bm \Sigma$ of one trajectory with a non-random initial condition $\z_0$. The difference is most pronounced for small observation intervals when the covariance of the ensemble $\P_f$ will be close to the initial analysis covariance $\P_a$, whereas a single trajectory will not have acquired much variance $\bm \Sigma$. In the long-time limit, both, $\P_f$ and $\bm \Sigma$, will approach the climatic covariance $\bm \Sigma_{\rm clim}$ (cf. (\ref{e.varOU})).\\
%Note that whereas ${\bm \Sigma}(\Delta \tobs)$ approaches $\bm \Sigma_{\rm clim}$ from below for $\Delta \tobs \to \infty$, $\P_f(t+\Delta \tobs)$ approaches the climatic state $\bm \Sigma_{\rm clim}$ from above.
In the following we restrict ourselves to the limit of small observation intervals $\Delta \tobs\ll 1$. In this limit, we can approximate $\P_a(t_i)\approx \P_f(t_{i+1})$ and explicitly solve the forecast covariance matrix $\P_f$ using (\ref{e.PfOU}). This assumption requires that the analysis is stationary in the sense that the filter has lost its memory of its initial background covariance provided by the user to start up the analysis. We have verified the validity of this assumption for small observation intervals and for a range of initial background variances. 
%Note that in standard ETKF this assumption becomes exact for all observation intervals in the limit $\Robs \to \infty$.
This assumption renders (\ref{e.PfOU}) a matrix equation for $\P_f$. To derive analytical expressions we further Taylor-expand the propagator $\L(\Delta \tobs)$ and the covariance $\bm \Sigma(\Delta \tobs)$ for small observation intervals $\Delta \tobs$. This is consistent with our stationarity assumption $\P_a(t_i)\approx \P_f(t_{i+1})$. The very lengthy analytical expression for $\P_f(t_{i+1})$ can be obtained with the aid of {\it Mathematica} \citep{Mathematica}, but is omitted from this paper.\\ In filtering one often uses variance inflation \citep{AndersonAnderson99} to compensate for the loss of ensemble variance due to finite size effects, sampling errors and the effects of nonlinearities. We do so here by introducing an inflation factor $\delta>1$ multiplying the forecast variance $\P_f$. Having determined the forecast covariance matrix $\P_f$ we are now able to write down an expression for the error covariance of the pseudo-observables $\Rw$. As before we limit the variance and the mean of our pseudo-observable $\y$ to be $\atarget = \bm {\sigma}_{\rm clim}^2$ and $\amtar = \bm \mu_{\rm clim}$. Then, upon using the definitions (\ref{e.BI}) and  (\ref{e.Rw}), we find that the error covariance for the pseudo-observables $\Rw$ is positive definite provided the observation interval $\Delta \tobs$ is sufficiently large\footnote{We actually compute $\Rw^{-1}$, however, since $\Rw$ is diagonal for our choice of the matrices, positive definiteness of $\Rw^{-1}$ implies positive definiteness of $\Rw$.}. Particularly, in the limit of $\Robs\to \infty$, we find that if
\begin{align}
\label{e.dtobsRwdelta}
\Delta \tobs(\delta) > \frac{\delta \lambda^2+4\gamma_x\gamma_y(1-\delta)}{2\gamma_x(1+\gamma_y^2)}\; ,
\end{align}
the variance constraint will be switched on. Note that for $\delta>1$ the critical $\Delta \tobs$ above which $\Rw$ is positive definite can be negative, implying that the variance constraint will be switched on for all (positive) values of $\Delta \tobs$. If no inflation is applied, i.e. $\delta=1$, this simplifies to
\begin{align}
\label{e.dtobsRw}
\Delta \tobs > \frac{\lambda^2}{2\gamma_x(1+\gamma_y^2)}>0\; . 
\end{align}
Because $4\gamma_x\gamma_y-\lambda^2>0$ the critical observation interval $\Delta \tobs$ is smaller for non-trivial inflation with $\delta>1$ than if no variance inflation is incorporated. This is intuitive, because the variance inflation will increase instances with $|\h\P_a\h^T|>|\bm {\sigma}_{\rm clim}^2|$. We have numerically verified that inflation is beneficial for the variance constraint to be switched on. It is pertinent to mention that for sufficiently large coupling strength $\lambda$ or sufficiently small values of $\gamma_x$, Equation (\ref{e.dtobsRw}) may not be consistent with the assumption of small observation intervals $\Delta \tobs\ll 1$.\\ We have checked analytically that the derivative of $\Rw^{-1}$ is positive at the critical observation interval $\Delta \tobs$, indicating that the frequency of occurrence when the variance constraint is switched on increases monotonically with the observation interval $\Delta \tobs$, in the limit of small $\Delta \tobs$. This has been verified numerically with the application of VLKF for (\ref{e.toy}) and is illustrated in Figure~\ref{fig:dtobstoymodel}.\\ At this stage it is important to mention effects due to finite size ensembles. For large observation intervals $\Delta \tobs \to \infty$ and large observational noise $\Robs \to \infty$, we have $\P_f \to \bm \Sigma_{\rm clim}$ and our analytical formulae would indicate that the variance constraint should not be switched on (cf. (\ref{e.BI}) and (\ref{e.Rw})). However, in numerical simulations of the Kalman filter we observe that for large observation intervals the variance constraint is switched on for almost all analysis times. This is a finite ensemble size effect and is due to the mean of the forecast variance ensemble adopting values larger than the climatic value of ${\bm \sigma}_{\rm clim}$ implying positive definite values of $\Rw$. The closer the ensemble mean approaches the climatic variance, the more likely fluctuations will push the forecast covariance above the climatic value. However, we observe that the actual eigenvalues of $\Rw$ decrease for $\Delta \tobs \to \infty$ and for the size of the ensemble $k\to \infty$.\\

\begin{figure}[htbp]
\begin{center}
\includegraphics[width = 0.5\linewidth,height=6cm]{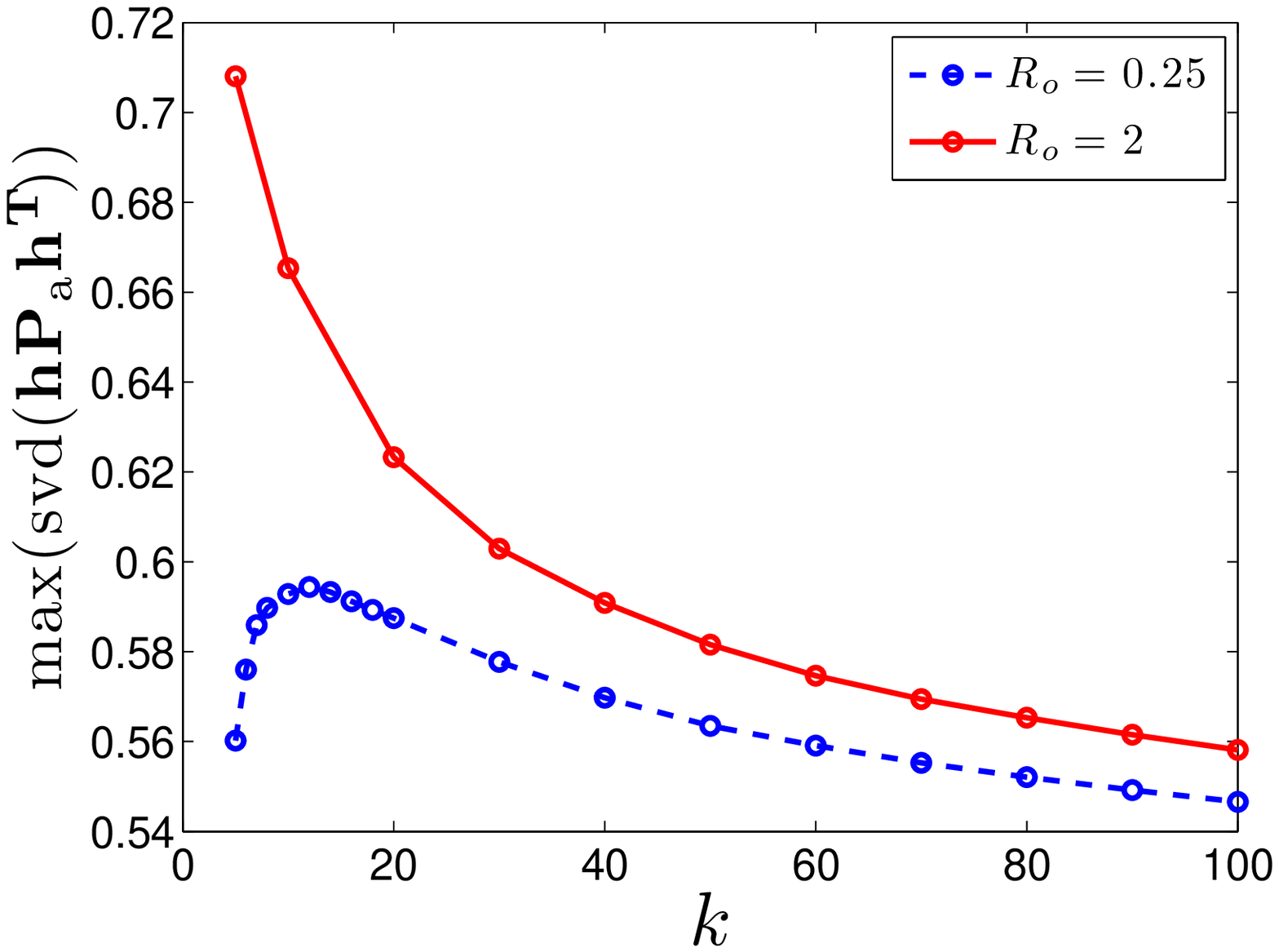}
\caption{Average maximal singular value of $\h \P_a \h^T$ as a function of ensemble size $k$ for the stochastic linear toy model (\ref{e.toy}) using standard ETKF without inflation, with $\Robs = 0.25$ (dashed curve) and $\Robs = 2$ (solid curve). Parameters are $\sigma_x = \sigma_y = \gamma_x = \gamma_y = 1$, $\lambda = 0.2$, $\Delta \tobs = 1$, for which the climatic variance is $\h \Sigma\h^T\approx0.505$. We used $50$ realizations for the averaging.
}
\label{fig:toy_hPahT}
\end{center}
\end{figure}

The analytical results obtained above are for the ideal case with $k\to \infty$. As mentioned in the introduction, in sparse observation networks finite ensemble sizes cause the overestimation of error covariances  \citep{Liuetal08,Whitakeretal09}, implying that $\Rw$ is positive definite and the variance limiting constraint will be switched on. This finite size effect is illustrated in Figure~\ref{fig:toy_hPahT}, where the maximal singular value of $\h \P_a \h^T$, averaged over $50$ realizations, is shown for ETKF as a function of ensemble size $k$ for different observational noise variances. Here we used no inflation, i.e. $\delta=1$, in order to focus on the effect of finite ensemble sizes. It is clearly seen that the projected covariance decreases for large enough ensemble sizes. The variance will asymptote from above to $\h {\bm \Sigma_{\rm clim}} \h^T$ in the limit $k\to \infty$. For sufficiently small observational noise, the filter corrects too large forecast error covariances by incorporating the observations into the analysis leading to a decrease in the analysis error covariance.\\

%
%%%%%%%%%%%%%%%%%%%%%%%%%%
%% programme used: toy_model_RwI_20100107.nb
%%%%%%%%%%%%%%%%%%%%%%%%%%

%

\noindent
However, the fact that the variance constraint is switched on does not necessarily imply that the variance limiting filter will perform better than the standard ETKF. In particular, for very large observation intervals $\Delta \tobs$ when the ensemble will have acquired the climatic mean and covariances, VLKF and ETKF will have equal skill. We now turn to the question under what conditions VLKF is expected to yield improved skill compared to standard ETKF. To this end we introduce as skill indicator the (squared) RMS error  
\begin{equation}
\label{e.rmslin}
{\cal{E}} =
\mathbb E^{t,dW} \|{\bar{\z}}_a(t_i) - \z_{\rm{truth}}(t_i) \|_{\bf G}^2\; ,
\end{equation}
between the truth $\z_{\rm{truth}}$ and the ensemble mean analysis ${\bar{\z}}_a$ (the square root is left out here for convenience of exposition). Here $\mathbb E^t$ denotes the temporal average over analyzes cycles, and $\mathbb E^{dW}$ denotes averaging over different realizations of the Brownian paths $dW$. We introduced the norm $\|{\bf{a}}{\bf{b}}\|_{\bf G}={\bf{a}}^T{\bf{G}}{\bf{b}}$ to investigate the overall skill using $\bf{G}=\Id$, the skill of the observed variables using ${\bf{G}}=\H^T\H$ and the skill of the pseudo-observables using ${\bf{G}}=\h^T\h$. Using the Kalman filter equation (\ref{e.zaens}) for the analysis mean with $\Kr=0$, we obtain for the ETKF
\begin{align*}
{\cal{E}}^{\rm ETKF} =
\mathbb E^{t,dW} \|(\Id-\KR \H)({\bar{\z}}_f(t_i) - \z_{\rm{truth}}(t_i)) + \KR \robs(t_i) \|_{\bf G}^2\; .
\end{align*}
Solving the linear toy-model (\ref{e.toy}) for each member of the ensemble and then performing an ensemble average, we obtain
\begin{align}
{\bar{\z}}_f(t_i)=\L(\Delta \tobs)\;{\bar{\z}}_a(t_{i-1})\; .
\end{align}
Substituting a particular realization of the truth $\z_{\rm{truth}}(t)$, and performing the average over the realizations, we finally arrive at
\begin{align}
\label{e_BW_col.epsETKF}
{\cal{E}}^{\rm ETKF} =
\mathbb E^{t} \|(\Id-\KR \H)\L(\Delta \tobs)\; \bm \xi_{t_{i-1}} \|_{\bf G}^2
+ 
\mathbb E^{t} \|(\Id-\KR \H)\bm \eta_{t_i} \|_{\bf G}^2
+
\mathbb E^{t} \|\KR \robs \|_{\bf G}^2
\; ,
\end{align}
with the mutually independent normally distributed random variables
\begin{eqnarray}
\bm \xi_{t_i} &=&{\bar{\z}}_a(t_i) -  \z_{\rm{truth}}(t_{i}) \sim {\cal{N}}(0,\P_a(t_{i}))
\nonumber
\\
\bm \eta_{t_i} &=& {\bm \sigma} \int_{t_{i-1}}^{t_i} \L(\Delta \tobs-s)\, d{\bm W}_s \sim {\cal{N}}(0,\bm \Sigma(\Delta \tobs))
\nonumber
\\
\robs &\sim& {\cal{N}}(0,\Robs) \; .
\end{eqnarray}
We have numerically verified the validity of our assumptions of the statistics of $\bm \xi_{t_i}$ and $\bm\eta_{t_i}$. Note that for $\bm \xi_{t_i}$ to have mean zero and variance $\P_a(t_{i})$ filter divergence has to be excluded. Similarly we obtain for the VLKF
\begin{eqnarray}
\label{e_BW_col.epsVLKFfull}
{\cal{E}}^{\rm VLKF} &=&
\mathbb E^{t} \|(\Id-\KR \H)\L(\Delta \tobs)\; \bm \xi_{t_{i-1}} \|_{\bf G}^2
+ 
\mathbb E^{t} \|(\Id-\KR \H)\bm \eta_{t_i} \|_{\bf G}^2
+
\mathbb E^{t} \|\KR \robs \|_{\bf G}^2
\nonumber
\\
&+&
\mathbb E^{t} \|\Kr \h \bm \zeta_{t_i} \|_{\bf G}^2
+2\, \mathbb E^{t} [(\Id-\KR \H)\L(\Delta \tobs)\; \bm \xi_{t_{i-1}})^T\, 
{\bf{G}}\, (\Kr \h \; \bm \zeta_{t_{i}})]
\; ,
\end{eqnarray}
with the normally distributed random variable
\begin{eqnarray}
\bm \zeta_{t_i} &=&{\bar{\z}}_f(t_i) \sim {\cal{N}}(0,\frac{1}{k}\P_f(t_i))
\; ,
\end{eqnarray}
where we used that $\amtar =0$. Note that using our stationarity assumption to calculate $\P_f$ we have $\zeta_{t_i} \, {\buildrel d \over \sim}\, (1/k) \xi_{t_{i-1}}$. Again we have numerically verified the statistics for $\bm \zeta_{t_i}$. The expression for the RMS error of the VLKF (\ref{e_BW_col.epsVLKFfull}) can be considerably simplified. Since for large ensemble sizes $k\to \infty$ the random variable $\zeta_{t_i}$ becomes a deterministic variable with mean zero, we may neglect all terms containing $\zeta_{t_i}$. We summarize to
\begin{eqnarray}
\label{e_BW_col.epsVLKF}
{\cal{E}}^{\rm VLKF} =
\mathbb E^{t} \|(\Id-\KR \H)\L(\Delta \tobs)\; \bm \xi_{t_{i-1}} \|_{\bf G}^2
+ 
\mathbb E^{t} \|(\Id-\KR \H)\bm \eta_{t_i} \|_{\bf G}^2
+
\mathbb E^{t} \|\KR \robs \|_{\bf G}^2 \; .
\end{eqnarray}
For convenience we have omitted superscripts for $\KR$ and $\bm \xi_{t_{i-1}}$ in (\ref{e_BW_col.epsETKF}) and (\ref{e_BW_col.epsVLKF}) to denote whether they have been evaluated for ETKF and VLKF. But note that, although the expressions (\ref{e_BW_col.epsETKF}) and (\ref{e_BW_col.epsVLKF}) are formally the same, one generally has ${\cal{E}}^{\rm ETKF}\neq  {\cal{E}}^{\rm VLKF}$, because the analysis covariance matrices $\P_a$ are calculated differently for both methods leading to different gain matrices $\KR$ and different statistics of $\bm \xi_t$ in (\ref{e_BW_col.epsETKF}) and (\ref{e_BW_col.epsVLKF}).

We can now estimate the skill improvement defined as 
\[
{\cal{S}}={\cal{E}}^{\rm ETKF}/{\cal{E}}^{\rm VLKF}
\]
 with values of ${\cal{S}}>1$ indicating skill improvement of VLKF over ETKF. We shall choose ${\bf{G}}=\h^T\h$ from now on, and concentrate on the skill improvement for the pseudo-observables. Recalling that ${\cal{E}}^{\rm ETKF}\approx {\cal{E}}^{\rm VLKF}$ for large observation intervals $\Delta \tobs$, we expect skill improvement for small $\Delta \tobs$. We perform again a Taylor expansion in small $\Delta \tobs$ of the skill improvement ${\cal{S}}$. The resulting analytical expressions are very lengthy and cumbersome, and are therefore omitted for convenience.\\ We found that there is indeed skill improvement ${\cal{S}}>1$ in the limit of either $\gamma_y\to \infty$ or $\gamma_x\to 0$.
%\note{The equations are horrible. We have tested that $\lim_{\gamma_y\to\infty} {\cal{S}}^{\Delta \tobs}=\infty$ and $\lim_{\gamma_x\to 0} {\cal{S}}^{\Delta \tobs}=\infty$, whereas $\lim_{\gamma_y\to 0}$ and $\lim_{\gamma_x\to \infty}$ are both finite. Here ${\cal{S}}^{\Delta \tobs}$ is ${\cal{S}}$ at first order in $\Delta \tobs$. Furthermore we have tested that $\partial {\cal{S}}/\partial_{\gamma_y}>0$, but also $\partial {\cal{S}}/\partial_{\gamma_x}>0$ at $\Delta \tobs=0$ but it can change sign. Note also that the dynamically effective order of magnitude of $\gamma_x$ is related to $\R_{\rm obs}$.  Shall we make the formulae more explicit in the text?}
This suggests that the skill is controlled by the ratio of the time scales of the observed and the unobserved variables. If the time scale of the pseudo-observables is much larger than the one of the observed variables, VLKF will exhibit superior performance over ETKF. This can be intuitively understood since $1/(2 \gamma_y)$ is the time scale on which equilibrium -- i.e. the climatic state --  is reached for the pseudo-observables $\y$. If the pseudo-observables have relaxed towards equilibrium within the observation interval $\Delta \tobs$, and their variance has acquired the climatic covariance $\h\P_a\h^T=\bm {\sigma}_{\rm clim}^2$, we expect the variance limiting to be beneficial.\\ 

Furthermore, we found analytically that the skill improvement increases with increasing observational noise $R_{\rm obs}$ (at least in the small observation interval approximation). In particular we found that $\partial {\cal{S}}/\partial R_{\rm obs}>0$ at $R_{\rm obs}=0$. The increase of skill with increasing observational noise can be understood phenomenologically in the following way. For $R_{\rm obs} = 0$ the filter trusts the observations, which as a time series carry the climatic covariance. This implies that there is a realization of the Wiener process such that the analysis can be reproduced by a model with the true values of $\gamma_{x,y}$ and $\sigma_{x,y}$. Similarly, this is the case in the other extreme $R_{\rm obs}\to \infty$, where the filter trusts the model. For $0\ll R_{\rm obs}\ll\infty$ the analysis reproducing system would have a larger covariance $\sigma_x$ than the true value. This slowed down relaxation towards equilibrium of the observed variables can be interpreted as an effective decrease of the damping coefficient $\gamma_x$. This effectively increases the time scale separation between the observed and the unobserved variables, which was conjectured above to be beneficial for skill improvement.\\

As expected, the skill improves with increasing inflation factor $\delta>1$. The improvement is exactly linear for $\Delta \tobs\to 0$. This is due to the variance inflation leading to an increase of instances with $\h\P_a\h^T>\bm {\sigma}_{\rm clim}^2$, for which the variance constraint will be switched on.\\

\begin{figure}[htbp]
\begin{center}
\includegraphics[width = 0.45\linewidth]{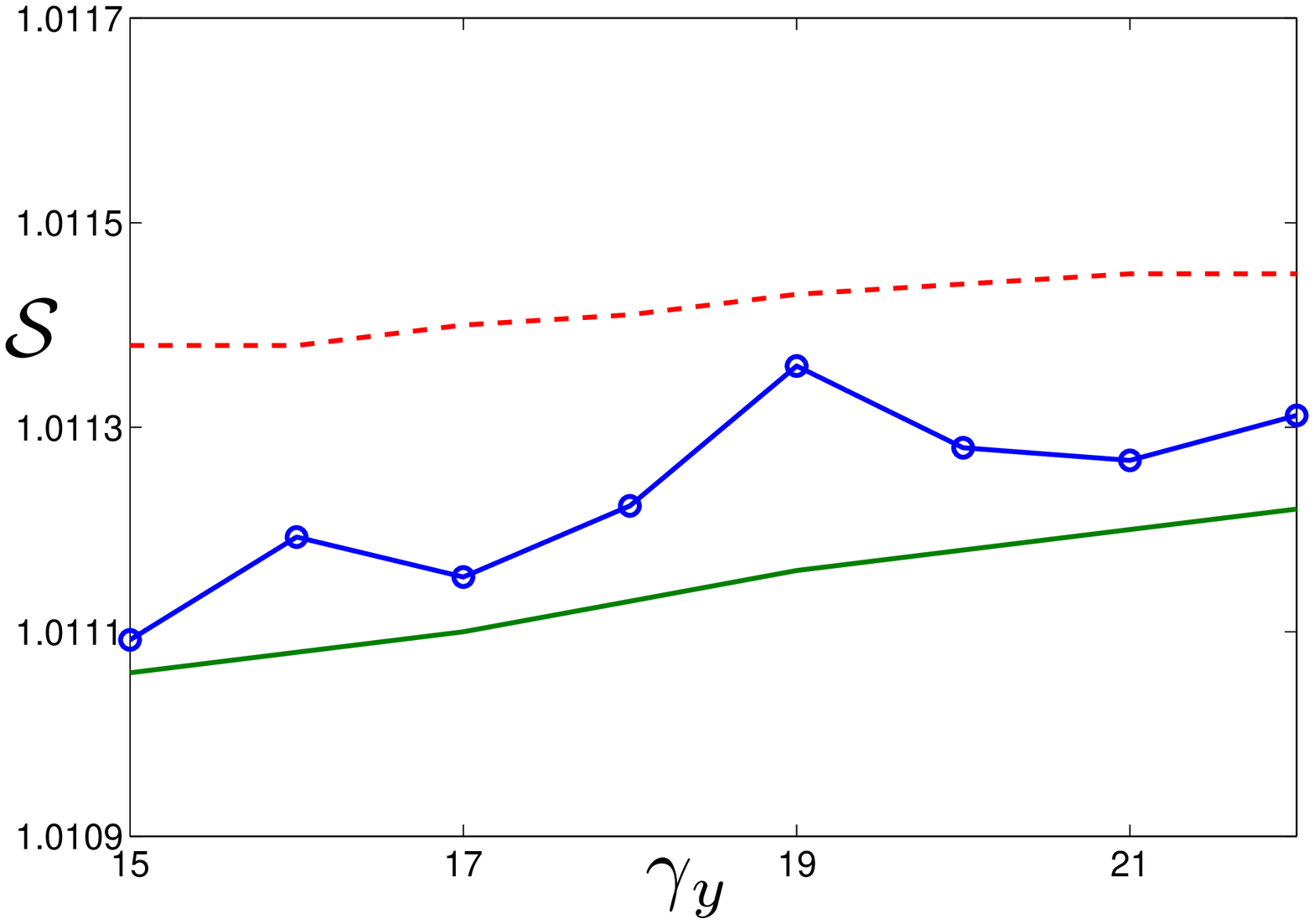}\quad
\includegraphics[width = 0.45\linewidth]{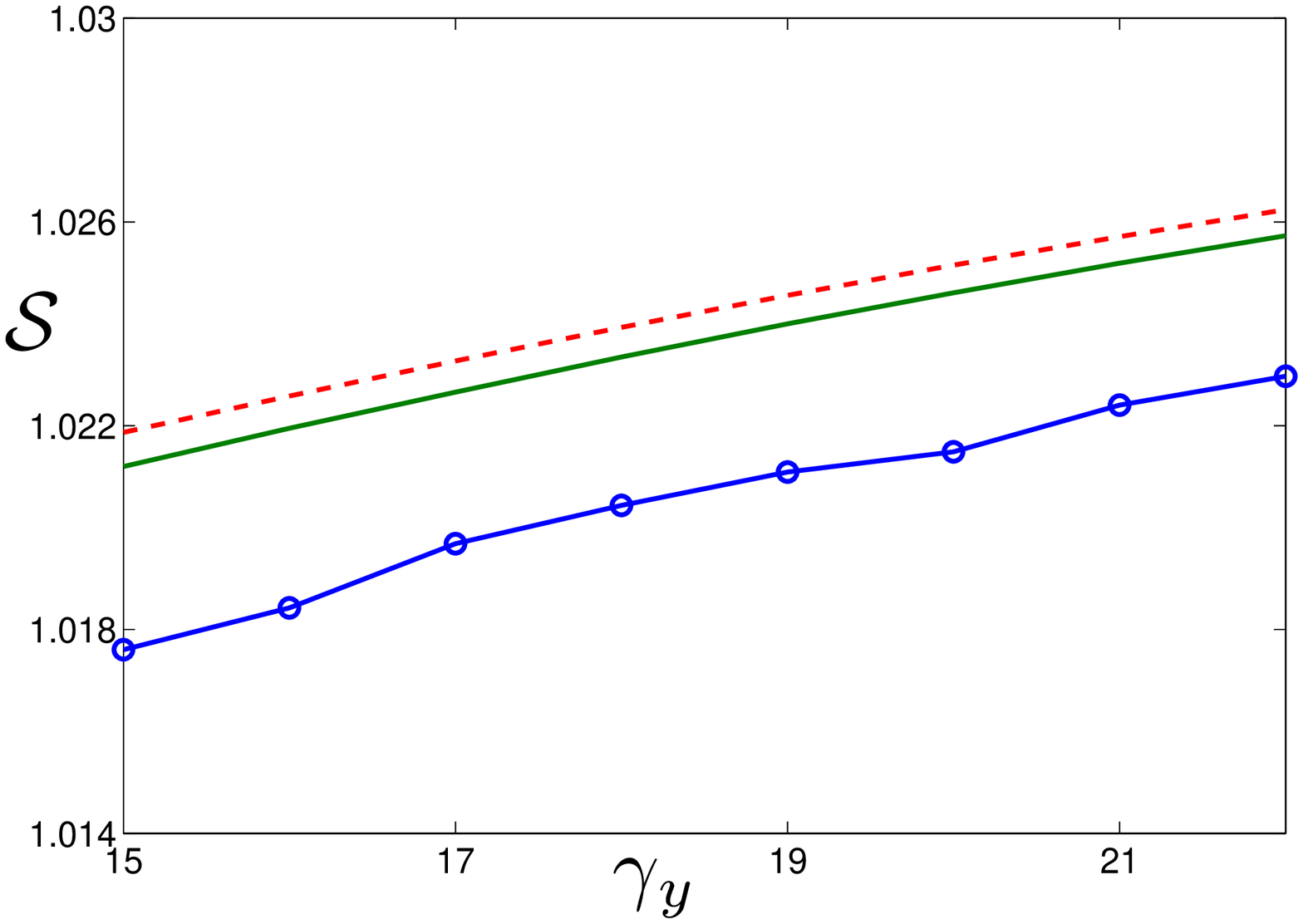}
\caption{Dependency of the skill improvement ${\cal{S}}$ of VLKF over ETKF on the damping coefficient $\gamma_y$ of the pseudo-observable. We show a comparison of direct numerical simulations (open circles)
%(online  blue
with analytical results using (\ref{e_BW_col.epsVLKFfull}) (continuous curve) 
%(online green)  
and the approximation of large ensemble size (\ref{e_BW_col.epsVLKF}) (dashed curve).
%, online red). 
Parameters are $\gamma_x=1$, $\lambda=2$, $\sigma_x=\sigma_y=1$, $R_{\rm obs}=0.25$. We used an ensemble size of $k=20$ and averaged over $1000$ realizations. Left: no inflation with $\delta=1$. Right: Inflation with $\delta=1.02^2$.}
\label{fig:skill_gy_infl}
\end{center}
\end{figure}

In Figure~\ref{fig:skill_gy_infl} we present a comparison of the analytical results (\ref{e_BW_col.epsETKF}) and (\ref{e_BW_col.epsVLKF}) with results from a numerical implementation of ETKF and VLKF for varying damping coefficient $\gamma_y$. Since $\gamma_y$ controls the time-scale of the $\y$-process, we cannot use the same $\Delta \tobs$ for a wide range of $\gamma_y$ in order not to violate the small observation interval approximations used in our analytical expressions. We choose $\Delta \tobs$ as a function of $\gamma_y$ such that the singular values of the first-order approximation of the forecast variance is a good approximation for this $\Delta \tobs$. For Figure~\ref{fig:skill_gy_infl} we have $\Delta \tobs\in(0.005,0.01)$ to preserve the validity of the Taylor expansion. Besides the increase of the skill with $\gamma_y$, Figure~\ref{fig:skill_gy_infl} shows that the value of ${\cal{S}}$ increases significantly for larger values of the inflation factor $\delta>1$.

%%%

%It is pertinent to mention that our results do not rely on the system involving a fast process.

We will see in the next Section that the results we obtained for the simple linear toy model (\ref{e.toy}) hold as well for a more complicated higher-dimensional model, where the dynamic Brownian driving noise is replaced by nonlinear chaotic dynamics. 

%

%%%%%%%%%%%%%%%%%%%%%%%%%%
%% programme used: skillETKFVLKF.nb
%%%%%%%%%%%%%%%%%%%%%%%%%%

%%%%%%%%%%%%%%%%%%%%%%%%%%%%%%%%%%%%%%%%%%

\section{Numerical results for the Lorenz-96 system}
\label{sec-numerics}
We illustrate our method with the Lorenz-96 system \citep{Lorenz96} and show its usefulness for sparse observations in improving the analysis skill and stabilizing the filter. In \citep{Lorenz96} Lorenz proposed the following model for the atmosphere
\begin{eqnarray}
\label{lorenz96}
{\dot{z}}_i=z_{i-1}(z_{i+1}-z_{i-2})-z_i+F
%+ \frac{h_x}{J}\sum_{j=1}^J y_{j,k}\\
%{\dot{y}}_{j,k}&=&\frac{1}{\epsilon}
%\left(y_{j+1,k}(y_{j+1,k}-x_{j+2,k})-y_{j,k}
%+ h_y x_k\right)\; ,
\qquad i=1,\cdots,D
\end{eqnarray}
with $\z=(z_1,\cdots,z_D)$ and periodic $z_{i+D}=z_i$. This system is a toy-model for midlatitude atmospheric dynamics, incorporating linear damping, forcing and nonlinear transport. The dynamical properties of the Lorenz-96 system have been investigated, for example, in \citep{LorenzEmanuel98,OrrellSmith03,Gottwaldmelbourne05}, and in the context of data assimilation it was investigated in, for example, \citep{Ott04,Fisher05,HarlimMajda09}. We use $D=40$ modes and set the forcing to $F=8$. These parameters correspond to a strongly chaotic regime \citep{Lorenz96}. For these parameters one unit of time corresponds to $5$ days in the earth's atmosphere as calculated by calibrating the $e$-folding time of the asymptotic growth rate of the most unstable mode with a time scale of $2.1$ days \citep{Lorenz96}. Assuming the length of a midlatitude belt to be about $30,000$km, the spatial scale corresponding to a discretization of the circumference of the earth along the midlatitudes in $D=40$ grid points corresponds to a spacing between adjacent grid points $z_i$ of approximately $750$km, roughly equalling the Rossby radius of deformation at midlatitudes. We estimated from simulations the advection velocity to be approximately $10.4$ m/sec which compares well with typical wind velocities in the midlatitudes.

In the following we will investigate the effect of using VLKF on improving the analysis skill when compared to a standard ensemble transform Kalman filter, and on stabilizing the filter and avoiding blow-up as discussed in \citep{Ott04,Kepert04,HarlimMajda09}. We perform twin experiments using a $k=41$-member ETKF and VLKF with the same truth time series, the same set of observations and the same initial ensemble. We have chosen an ensemble with $k>D$ in order to eliminate the effect that a finite-size ensemble can only fit as many observations as the number of its ensemble members \citep{Lorenc03}. Here we want to focus on the effect of limiting the variance.

The system is integrated using the implicit mid-point rule (see for example \cite{LeimkuhlerReich}) to a time $T=30$ with a time step $dt=1/240$. The total time of integration corresponds to an equivalent of $150$ days, and the integration timestep $dt$ corresponds to half an hour. We measured the approximate climatic mean and variance, $\muclim$ and  $\sigmaclim^2$, respectively, via a long time integration over a time interval of $T=2000$ which corresponds roughly to $27.5$ years. Because of the symmetry of the system (\ref{lorenz96}), the mean and the standard deviation are the same for all variables $z_i$, and are measured to be $\sigmaclim=3.63$ and $\muclim=2.34$.

The initial ensemble at $t=0$ is drawn from an ensemble with variance $\sigmaclim^2$; the filter was then subsequently spun up for sufficiently many analysis cycles to ensure statistical stationarity. We assume Gaussian observational noise of the order of $25$\% of the climatological standard deviation $\sigma_{clim}$, and set the observational error covariance matrix $\Robs=(0.25\sigma_{clim})^2  \,\mathbf{I}$. We find that for larger observational noise levels the variance limiting correction (\ref{e.Rw}) is used more frequently. This is in accordance with our finding in the previous section for the toy model.

%We have used a constant variance inflation factor $\delta=1.05$ for both filters. We note that the optimal inflation factor at which the RMS error $\cal{E}$ is minimal, is different for VLKF and ETKF. For $\delta t_{obs} = 0.05$ and $N_{skip} = 4$ we find that $\delta=1.06$ produces minimal RMS error for the VLKF and $\delta = 1.03$ produces minimal RMS error for ETKF, and for too small inflation factors $\delta<1.04$ we observed instances of filter divergencies. 

We study first the performance of the filter and its dependence on the time between observations $\Delta \tobs$ and the proportion of the system observed $1 /\Nobs$. $\Nobs=2$ means only every second variable is observed, $\Nobs=4$ only every fourth, and so on.

We have used a constant variance inflation factor $\delta=1.05$ for both filters. We note that the optimal inflation factor at which the RMS error $\cal{E}$ is minimal, is different for VLKF and ETKF. For $\Delta t_{obs} = 5/120$ ($5$ hours) and $N_{obs} = 4$ we find that $\delta=1.06$ produces minimal RMS errors for VLKF and $\delta = 1.04$ produces minimal RMS errors for ETKF. For $\delta<1.04$ filter divergence occurs in ETKF, so we chose $\delta=1.05$ as a compromise between controlling filter divergence and minimizing the RMS errors of the analysis. 

\begin{figure}[htbp]
\begin{center}
\includegraphics[width = \linewidth,height=8cm]{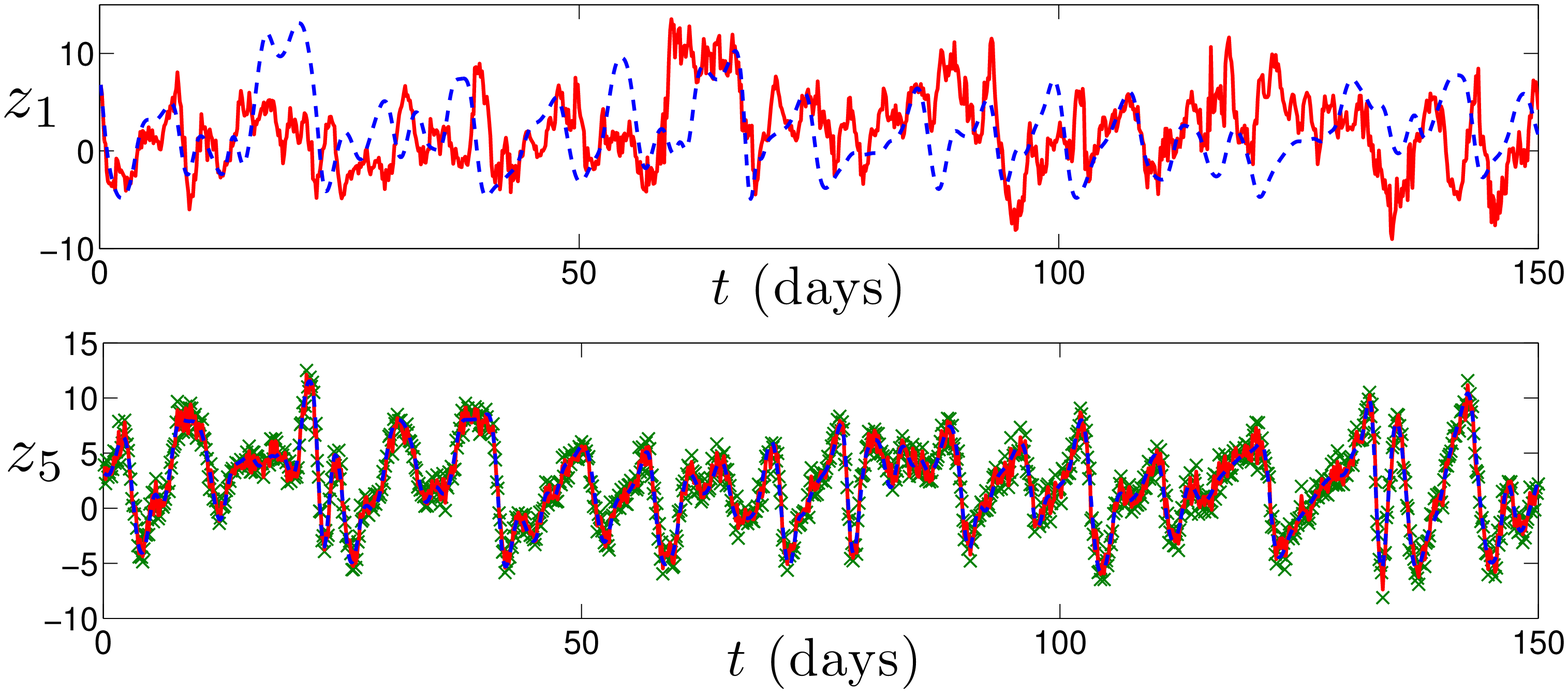}
\caption{Sample ETKF analysis (continuous grey line)
%(blue) 
for observed $z_5$ (bottom panel) and unobserved $z_1$ (top panel) component. 
%Red line 
The dashed line is the truth, 
%green 
the crosses are observations. Parameters used were  $N_{obs} = 5$, $\Delta \tobs = 0.15$ (18 hours) and $\Robs = (0.25\sigma_{clim})^2 \, \mathbf{I}$.}
\label{fig:ETKF_analysis}
\end{center}
\end{figure}

\begin{figure}[htbp]
\begin{center}
\includegraphics[width = \linewidth,height=8cm]{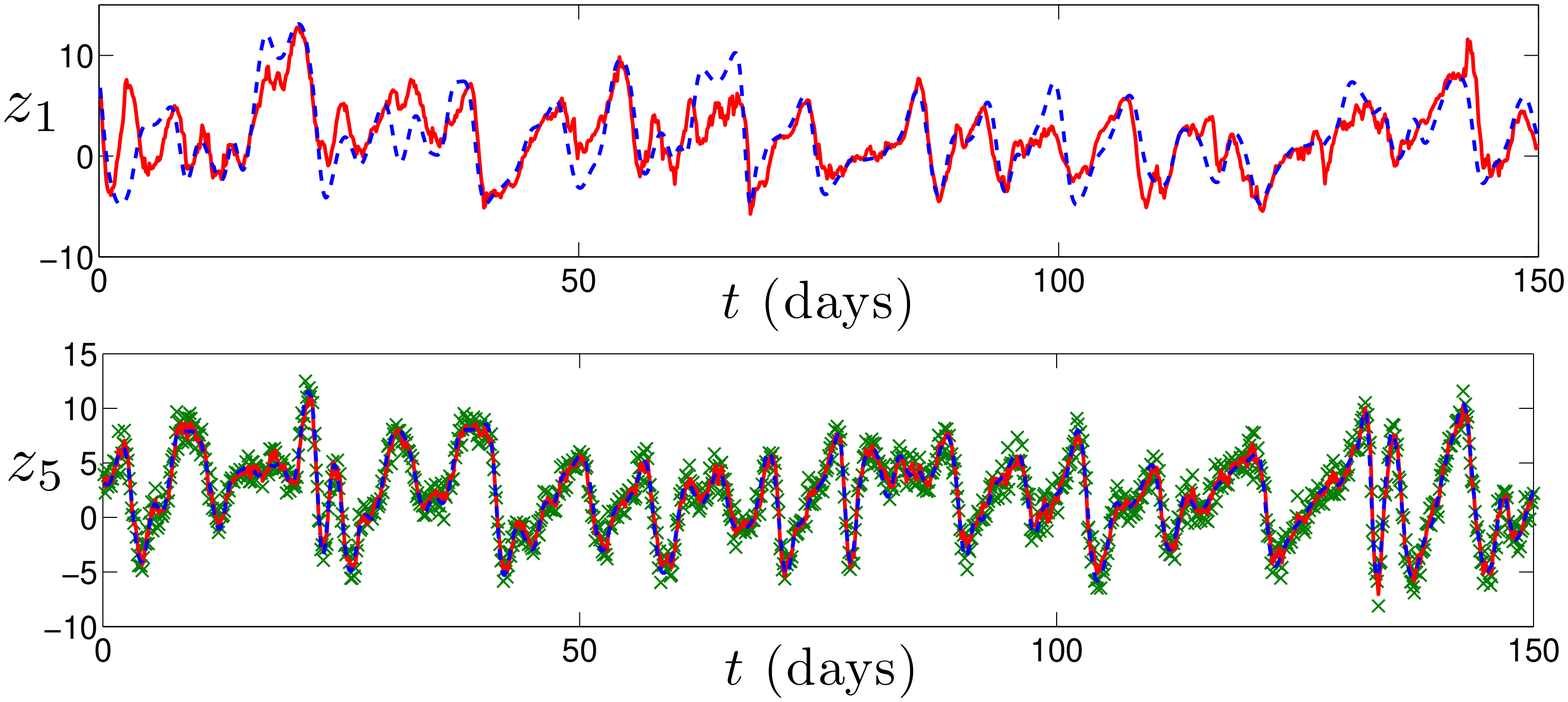}
\caption{Sample VLKF analysis (continuous grey line) for observed $z_5$ (bottom panel) and unobserved $z_1$ (top panel) component. 
%Red line 
The dashed line is the truth, 
%green 
the crosses are observations. Parameters as in Figure~\ref{fig:ETKF_analysis}.}
\label{fig:VLKF_analysis}
\end{center}
\end{figure}

Figure \ref{fig:ETKF_analysis} shows a sample analysis using ETKF with $N_{obs} = 5$, $\Delta \tobs = 0.15$ and $\Robs = (0.25\sigma_{clim})^2 \, \mathbf{I}$ for an arbitrary unobserved component (top panel) and an arbitrary observed component (bottom panel) of the Lorenz-96 model. While the figure shows that the analysis (continuous grey line) tracks the truth (dashed line) reasonably well for the observed component, the analysis is quite poor for the unobserved component. Substantial improvements are seen for the VLKF when we incorporate information about the variance of the un-observed pseudo-observables, as can be seen in Figure~\ref{fig:VLKF_analysis}. We set the mean and the variance of the pseudo-observables to be the climatic mean and variance, $\amtar = \muclim {\bf e}$ and $\atarget= \sigma_{\rm clim}^2 {\bf I}$ to filter the same truth with the same observations as used to produce Fig.~\ref{fig:ETKF_analysis}. For these parameters (and in this realization) the quality of the analysis in both the observed and unobserved components is improved.\\


\begin{figure}[htbp]
\begin{center}
\includegraphics[width = 0.5\linewidth,height=6cm]{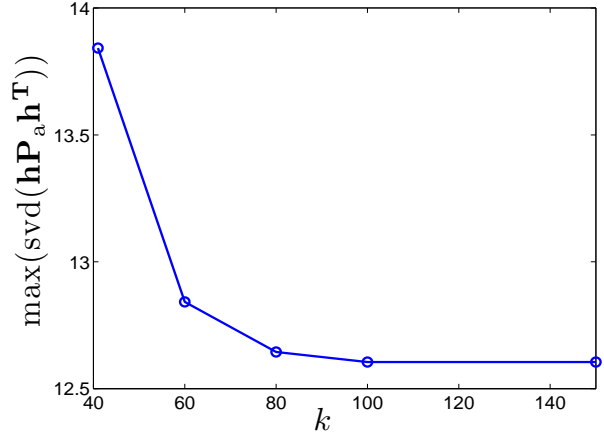}
\caption{Average maximal singular value of $\h \P_a \h^T$ as a function of ensemble size $k$ for the Lorenz-96 model (\ref{lorenz96}), using standard ETKF without inflation.  All other parameters are as in Figure~4. We used $150$ realisations for the averaging.}
\label{fig:L96_hPahT}
\end{center}
\end{figure}

As for the linear toy model (\ref{e.toy}), finite ensemble sizes exacerbate the overestimation of error covariances. In Figure~\ref{fig:L96_hPahT} the maximal singular value of $\h \P_a \h^T$, averaged over $150$ realizations, is shown for ETKF as a function of ensemble size $k$. Again we use no inflation, i.e. $\delta=1$, in order to focus on the effect of finite ensemble sizes. The projected covariance clearly decreases for large enough ensemble sizes. However, here the limit of the maximal singular value of $\h \P_a \h^T$ for $k\to\infty$ underestimates the climatic variance $\sigmaclim^2 = 13.18$.\\

\begin{figure}[htbp]
\begin{center}
\includegraphics[width = 0.5\linewidth]{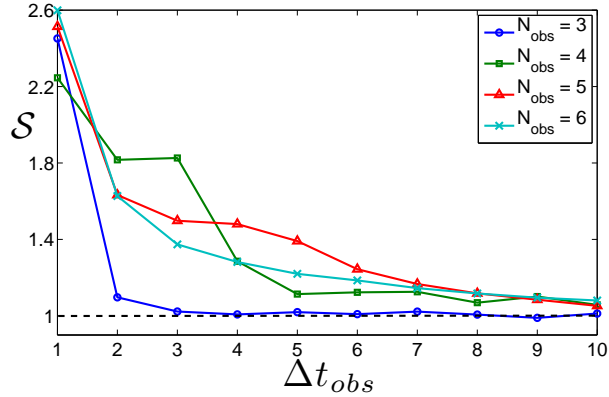}\\
\caption{Proportional skill improvement of VLKF over ETKF as a function of the observation interval $\Delta \tobs$ for different values of $\Nobs$, with observational noise $\Robs = (0.25\sigma_{clim})^2 \, \mathbf{I}$. A total of $500$ simulations were used to perform the ensemble average in the RMS errors ${\cal{E}}$ using (\ref{e.rms}) for ETKF and VLKF. $\Delta \tobs$ is measured in hours.}
\label{fig:VLKF_improvement}
\end{center}
\end{figure}

\noindent
To quantify the improvement of the VLKF filter we measure the site-averaged RMS error
\begin{equation}
\label{e.rms}
{\cal{E}} =
\sqrt{
\langle
\frac{1}{LD_o}\sum_{l=1}^L \|{\bar{\z}}_a(l \Delta \tobs) - \z_{\rm{truth}}(l \Delta \tobs) \|^2
\rangle
}
\end{equation}
between the truth $\z_{\rm{truth}}$ and the ensemble mean ${\bar{\z}}_a$ with $L=\lfloor T/\Delta \tobs \rfloor$ where the average is taken over $500$ different realizations, and $D_o\le D$ denotes the length of the vectors $
{\bar{\z}}_a$. In tables \ref{table:skillVLKF} we display ${\cal{E}}$ for the ETKF and VLKF respectively, as a function of $N_{obs}$ and $\Delta \tobs$. The increased RMS error for larger observation intervals $\Delta \tobs$ can be linked to the increased variance of the chaotic nonlinear dynamics generated during longer integration times between analyses.  Figure~\ref{fig:VLKF_improvement} shows the average proportional improvement of the VLKF over ETKF, obtained from the values of tables \ref{table:skillVLKF}. Figure~\ref{fig:VLKF_improvement} shows that the skill improvement is greatest when the system is observed frequently. For large observation intervals $\Delta \tobs$ ETKF and VLKF yield very similar RMS. We checked that for large observation intervals $\Delta \tobs$ both filters still produce tracking analyses. Note that the observation intervals $\Delta \tobs$ considered here are all much smaller than the $e$-folding time of $2.1$ days. The most significant improvement occurs when one quarter of the system is observed, that is for $N_{obs} = 4$, and for small observation intervals $\Delta \tobs$. The dependency of the skill of VLKF on the observation interval is consistent with our analytical findings in Section~\ref{sec-toymodel}. 

We have tested that the increase in skill as depicted in Figure~\ref{fig:VLKF_improvement} is not sensitive to incomplete knowledge of the statistical properties of the pseudo-observables by perturbing $\atarget$ and $\amtar$ and then monitoring the change in RMS error. We performed simulations where we drew $\atarget$ and $\amtar$ independently from uniform distributions $(0.9 \, \atarget , 1.1 \, \atarget)$ and $(0.9 \, \amtar, 1.1\, \amtar)$. We found that for parameters $N_{obs} = 2, 4, 6$, $\eta = 0.05, 0.25, 0.5$ (with $\eta$ measuring the amount of the climatic variance used through $\Robs = (\eta\,\sigma_{clim})^2 \, \mathbf{I}$), and $\Delta\tobs = 0.025, 0.05, 0.25$ (corresponding to $3$, $6$ and $30$ hours) over a number of simulations there was on average no more than 7\% difference of the analysis mean and the singular values of the covariance matrices between the control run where $\atarget= \sigma_{\rm clim}^2 {\bf I}$ and $\amtar=\muclim {\bf e}$ is used, and when $\atarget$ and $\amtar$ are simultaneously perturbed.

\begin{figure}[htbp]
\begin{center}
\includegraphics[width = 0.475\linewidth]{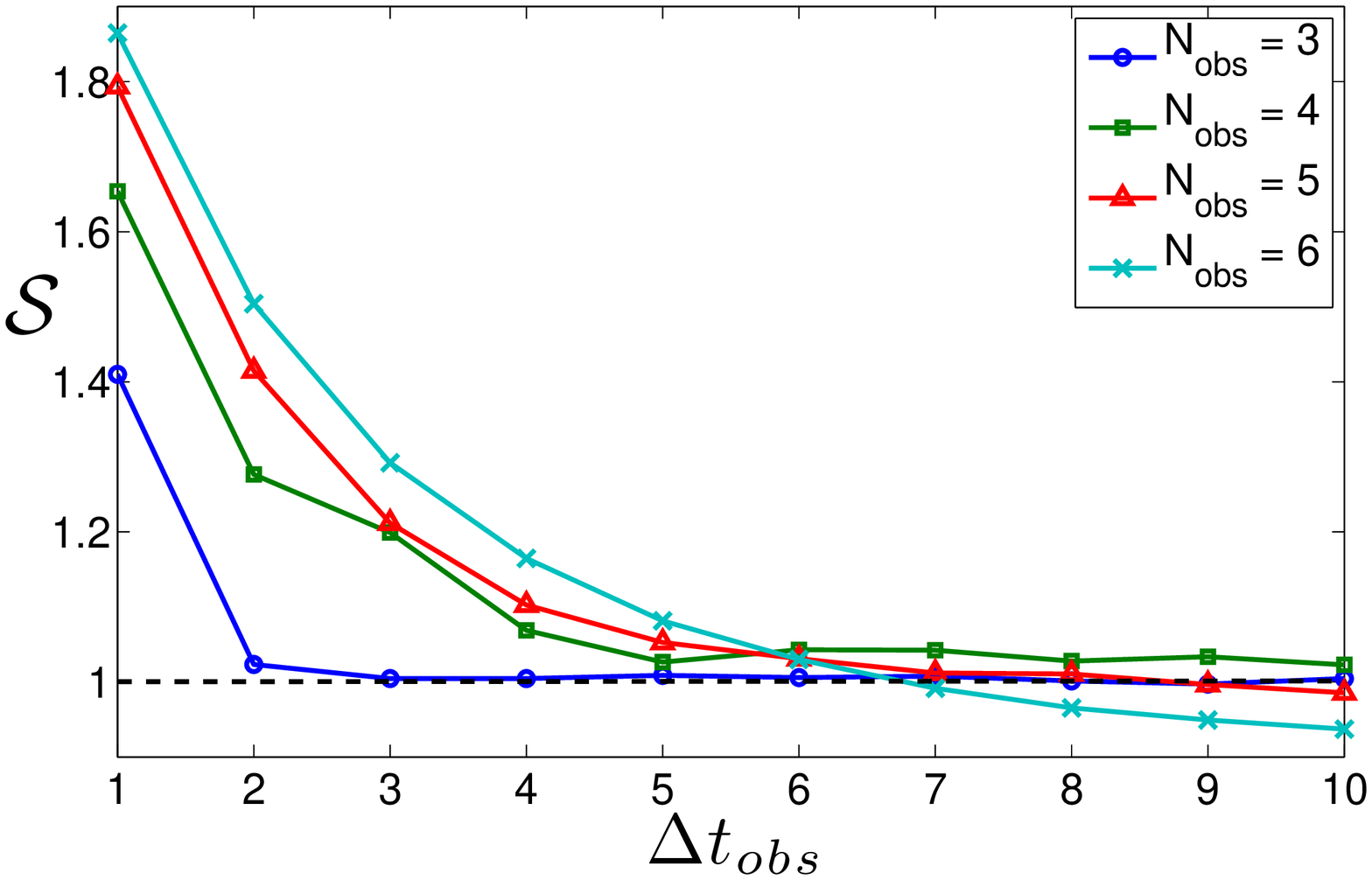}
\includegraphics[width = 0.475\linewidth]{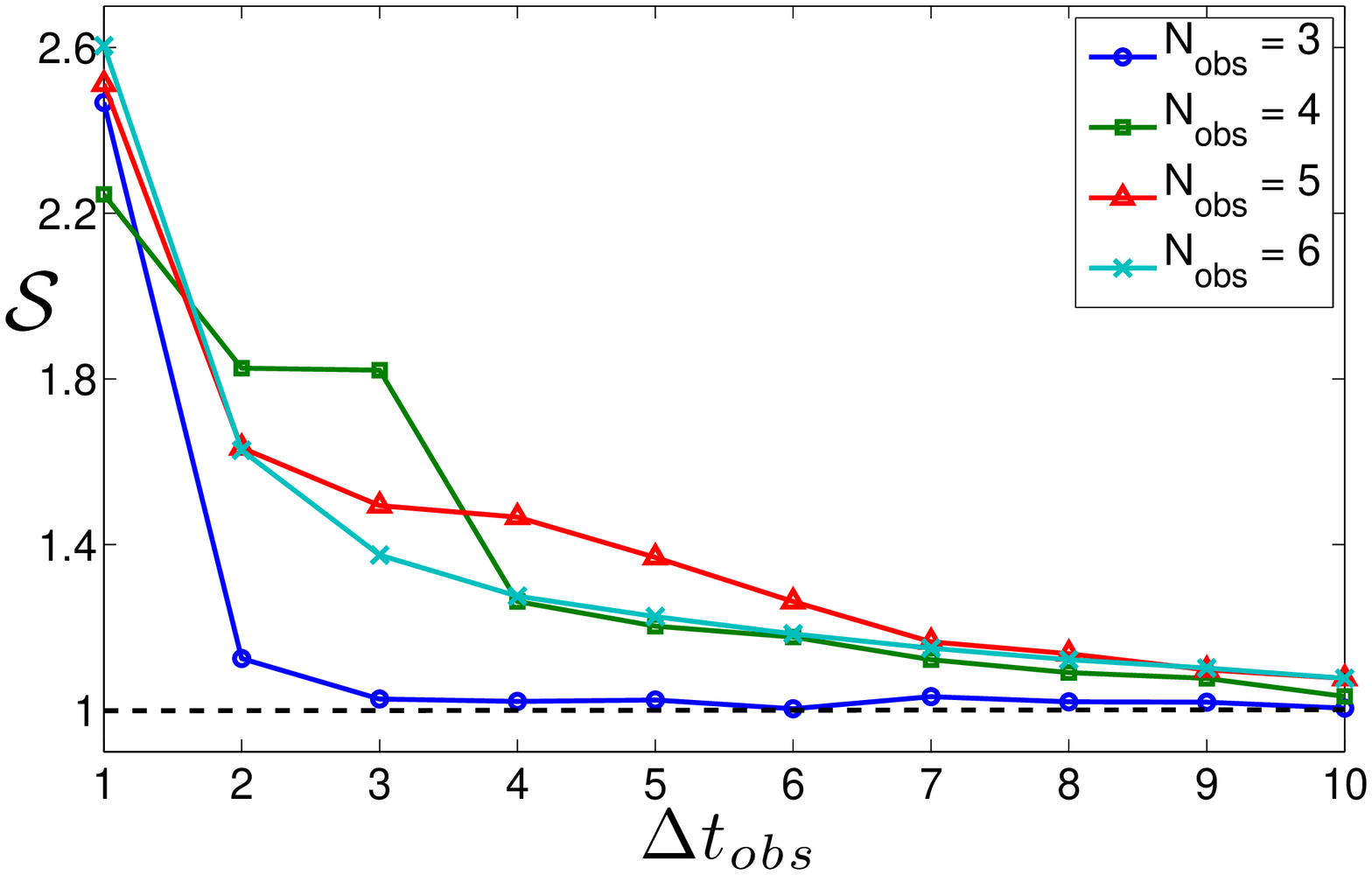}
\caption{Proportional skill improvement of VLKF over ETKF as a function of the observation interval $\Delta \tobs$ for different values of $\Nobs$. The RMS error ${\cal{E}}$ is calculated using only the observed variables (left) or only the pseudo-observables (right). $\Delta \tobs$ is measured in hours. Parameters as in Figure~\ref{fig:VLKF_improvement}.}
\label{fig:VLKF_improvement_2}
\end{center}
\end{figure}

\begin{figure}[htbp]
\begin{center}
\includegraphics[width = 0.475\linewidth]{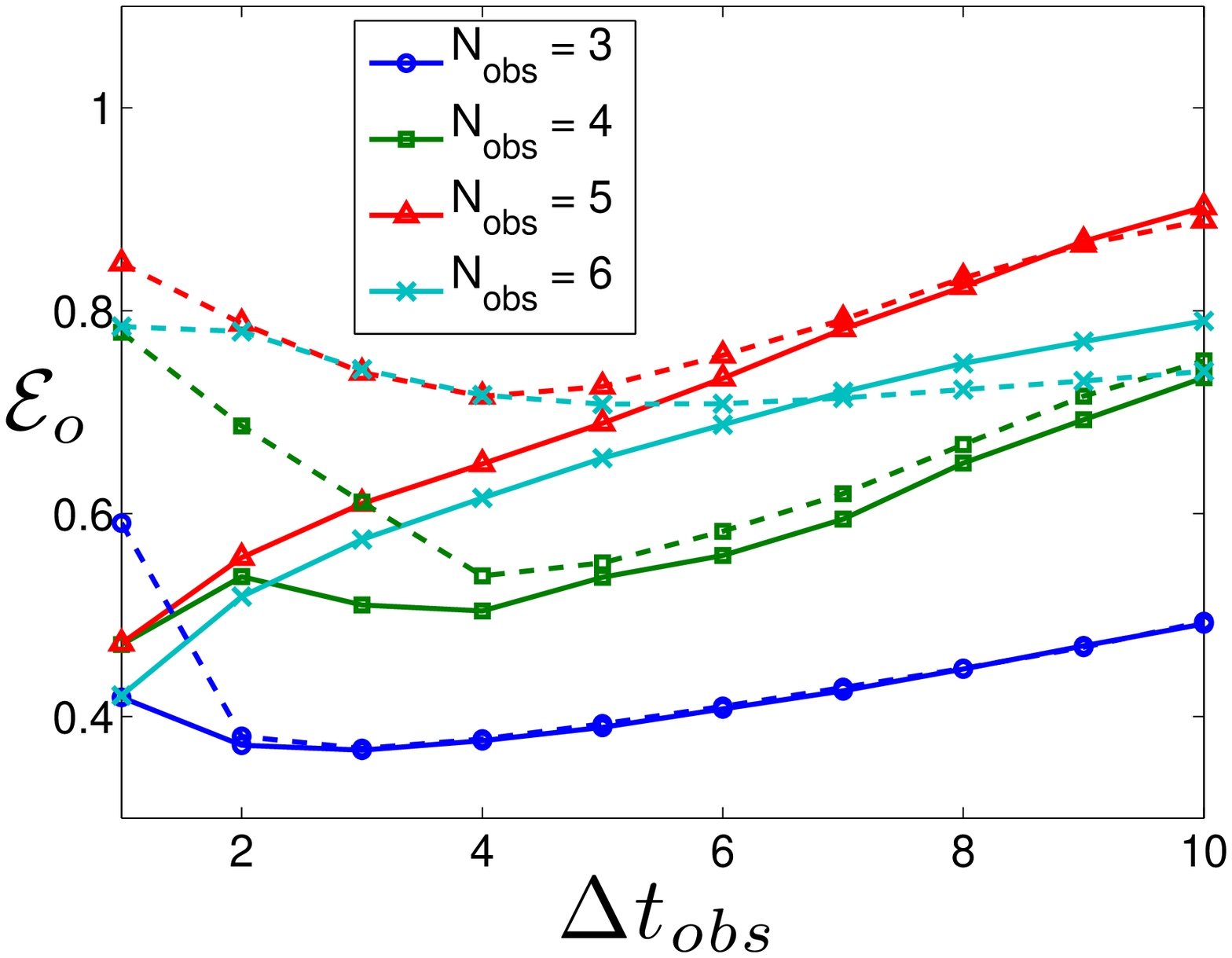}
\includegraphics[width = 0.475\linewidth]{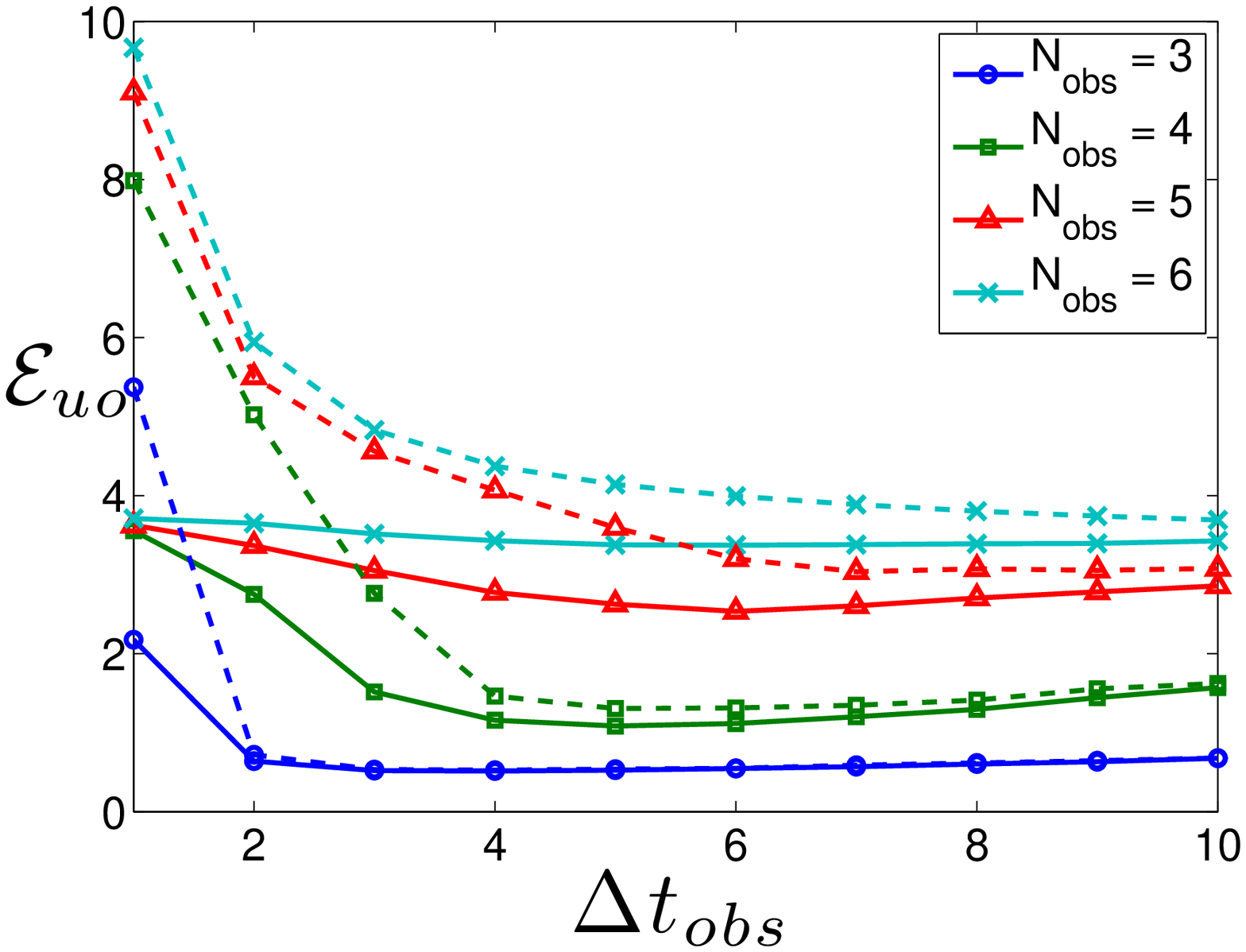}\caption{RMS error of VLKF (solid lines) and ETKF (dashed lines) for $\Robs = (0.25\sigma_{clim})^2 \, \mathbf{I}$, where ${\cal{E}}$ is calculated using only the observed variables (left) or only the pseudo-observables (right). $\Delta \tobs$ is measured in hours. Parameters as in Figure~\ref{fig:VLKF_improvement}.}
\label{fig:VLKF_improvement_3}
\end{center}
\end{figure}

\begin{figure}[htbp]
\begin{center}
\includegraphics[width = 0.5\linewidth]{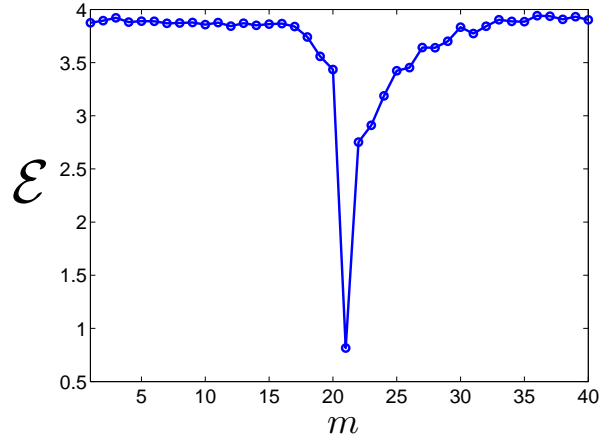}\\
\caption{RMS error ${\cal{E}}$ for each variable $z_i$ as a function of the lattice site $i$. Only one observable was used at $i=21$. Time between observations is $\Delta \tobs = 10$ hours and observational noise with covariance $\Robs=(0.25\, \sigma_{clim}^2)\Id$ was used. The results are averaged over $100$ different realizations.}
\label{fig:asymmetry}
\end{center}
\end{figure}

An interesting question is how the relative skill improvement is distributed over the observed and unobserved variables. This is illustrated in Figure~\ref{fig:VLKF_improvement_2} and Figure~\ref{fig:VLKF_improvement_3}. In Figure~\ref{fig:VLKF_improvement_2} we show the proportional skill improvement of VLKF over ETKF for the observed variables and the pseudo-observables, respectively. Figure~\ref{fig:VLKF_improvement_2} shows that the skill improvement is larger for the pseudo-observables than for the observables which is to be expected. In Figure~\ref{fig:VLKF_improvement_3} we show the actual RMS error ${\cal{E}}$ for ETKF and VLKF for the observed variables and the pseudo-observables. It is shown that the skill improvement is better for the unobserved pseudo-observables for all observation intervals $\Delta \tobs$. In contrast, VLKF exhibits an improved skill for the observed variables either for small observation intervals for all values of $\Nobs$ or for all (sufficiently small) observation intervals when $\Nobs=4, 5$. We have, however, checked that the analysis is still tracking the truth reasonably well, and the discrepancy with ETKF is not due to the analysis not tracking the truth anymore. As expected, the RMS error asymptotes for large observation intervals $\Delta \tobs$ (not shown) to the standard deviation of the observational noise $0.25\, \sigmaclim \approx 0.91$ for the observables, and to the climatic standard deviation $\sigmaclim=3.63$ for the pseudo-observable  (not shown), albeit slightly reduced for small values of $\Nobs$ due to the impact of the surrounding observed variables (see Figure~\ref{fig:asymmetry}).

Note that there is an order of magnitude difference between the RMS errors for the observables and the pseudo-observables  for large $\Nobs$ (cf. Figures~\ref{fig:VLKF_improvement_3}). This suggests that the information of the observed variables does not travel too far away from the observational sites. However, the nonlinear coupling in the Lorenz-96 system (\ref{lorenz96}) allows for information of the observed components to influence the error statistics of the unobserved components. Therefore the RMS error of pseudo-observables adjacent to observables are better than those far away from observables. Moreover, the specific structure of the nonlinearity introduces a translational symmetry-breaking (one may think of the nonlinearity as a finite difference approximation of an advection term $\z\z_x$), which causes those pseudo-observables to the right of an observable to have a more reduced RMS error than those to the left of an observable. This is illustrated in Figure~\ref{fig:asymmetry} where the RMS error is shown for each site when only one site is observed. The advective time scale of the Lorenz-96 system is much smaller than $\Delta \tobs$ which explains why the skill is not equally distributed over the sites, and why, especially for large values of $\Nobs$, we observe a big difference between the site-averaged skills of the observed and unobserved variables.

\begin{table}
\begin{center}
\begin{tabular}{|r r||c|c|c|c|c|c|c|c|c|c|}
 \hline
 \multirow{6}{*}{$N_{obs}$ }\vline & 6 & 4.40  &  3.64  &  3.42  &  3.32  &  3.29  &  3.30  &  3.30  &  3.28  &  3.26  &  3.26\\
\vline& 5 & 4.08  &  2.88  &  2.70  &  2.83  &  3.02  &  3.07  &  3.17  &  3.21  &  3.19  &  3.20\\
\vline& 4 & 2.42  &  1.17  &  1.35  &  1.72  &  2.18  &  2.37  &  2.62 & 2.84  &  2.98  &  3.06\\
\vline& 3 & 0.49  &  0.51  &  0.60 &   0.71  &  0.89  &  1.11  &  1.38  &  1.68  &  2.02  &  2.25\\
\vline& 2 & 0.31  &  0.34  &  0.38  &  0.43  &  0.49  &  0.55  &  0.66  &  0.75  &  0.90  &  1.13\\
\vline& 1 &    0.19 &   0.21  &  0.24  &  0.26  &  0.29  &  0.31  &  0.33  &  0.36  &  0.39  &  0.44\\
 \hline
 \hline
& & 0.025 & 0.05 & 0.075 & 0.1 & 0.125 & 0.15 & 0.175 & 0.2 & 0.225 & 0.25 \\
& & 3 h & 6 h & 9 h &  12 h & 15 h & 18 h & 21 h & 24 h & 27 h & 30 h \\
 \cline{3-12}
 & & \multicolumn{10}{|c|}{$\Delta \tobs$} \\
 \hline
%\end{tabular}
%\end{center}
%\label{table:skillETKF}
%
%\begin{center}
%\begin{tabular}{|r r||c|c|c|c|c|c|c|c|c|c|}
 \hline
 \multirow{6}{*}{$N_{obs}$ }\vline & 6 &  3.20  &  3.09  &  3.10  &  3.15  &  3.20  &  3.22  &  3.27  &  3.27  & 3.26  &  3.27\\
\vline& 5 & 2.73  &  2.28  &  2.51  &  2.70  &  2.89  &  3.03  &  3.07  &  3.14  &  3.15  &  3.15\\
\vline& 4 & 1.30  &  1.03  &  1.28  &  1.66  &  2.04  &  2.29  &  2.55  &  2.70  &  2.88  &  2.96\\
\vline& 3 & 0.48  &  0.51  &  0.59  &  0.70  &  0.87  &  1.07  &  1.39  &  1.71  &  1.95  &  2.21\\
\vline& 2 &  0.31  &  0.34 & 0.38  &  0.44  &  0.50  &  0.56  &  0.64  &  0.77  &  0.95  &  1.14\\
\vline& 1 & 0.19  &  0.21  &  0.24  &  0.26  &  0.29 &  0.31  &  0.33  &  0.36  &  0.39  &  0.44\\
 \hline
 \hline
& & 0.025 & 0.05 & 0.075 & 0.1 & 0.125 & 0.15 & 0.175 & 0.2 & 0.225 & 0.25 \\
& & 3 h & 6 h & 9 h &  12 h & 15 h & 18 h & 21 h & 24 h & 27 h & 30 h \\
 \cline{3-12}
 & & \multicolumn{10}{|c|}{$\Delta \tobs$} \\
 \hline
 \end{tabular}
\end{center}
\caption{RMS errors for ETKF (upper table) and VLKF (bottom table), averaged over 500 simulations, and with $\Robs = (0.25\sigma_{clim})^2 \, \mathbf{I}$ as observational noise.}
\label{table:skillVLKF}
\end{table}


\begin{figure}[htbp]
\begin{center}
\includegraphics[width = 0.5\linewidth]{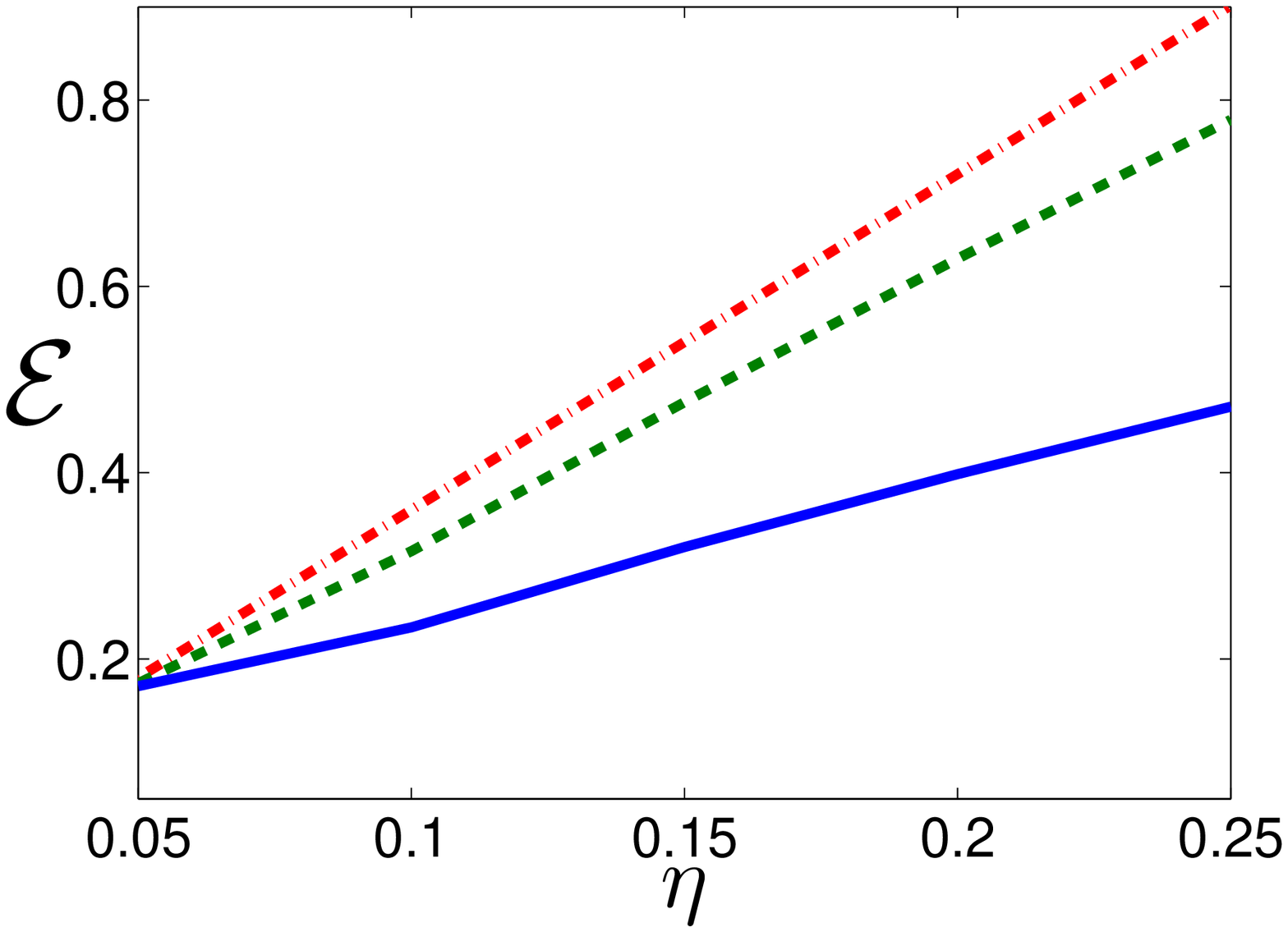}\\
\includegraphics[width = 0.5\linewidth]{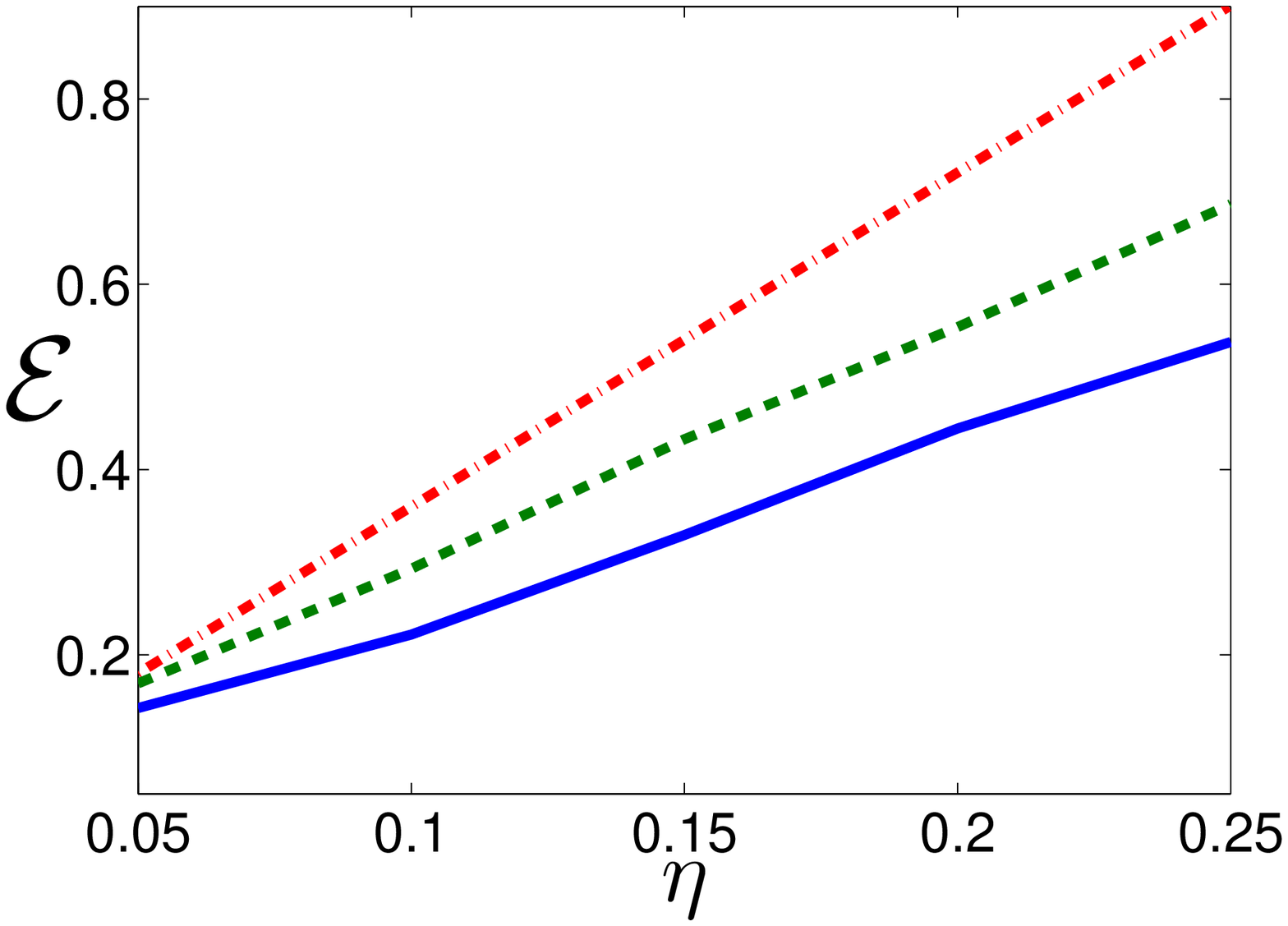}\\
\includegraphics[width = 0.5\linewidth]{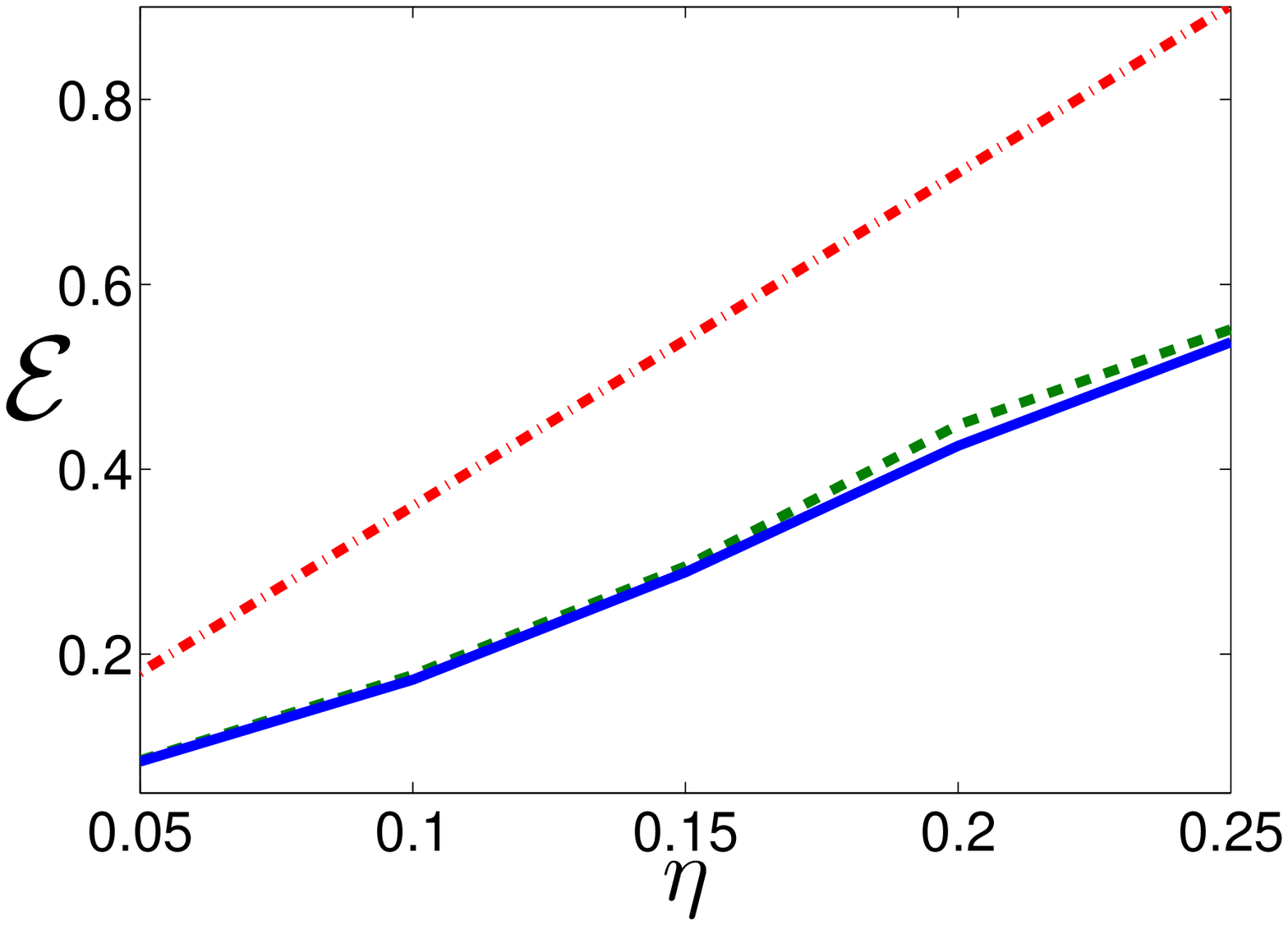}
\caption{RMS error $\cal{E}$ of VLKF (solid lines) and ETKF (dashed lines), as a function of the observational noise, measured here by $\eta$ defined via $\Robs = (\eta\,\sigma_{clim})^2 \, \mathbf{I}$. The dotted line indicates the RMS error if only observations were used. We show results from top to bottom for several observation intervals:  $\Delta \tobs=1$ hour, $\Delta \tobs=2$ hours and $\Delta \tobs=5$ hours. $\Nobs=4$ was used and $1000$ simulations were carried out to perform the ensemble averages in the RMS errors ${\cal{E}}$ using (\ref{e.rms}) for ETKF and VLKF.}
\label{fig:Rdep}
\end{center}
\end{figure}

In Figure~\ref{fig:Rdep} we show how the RMS error behaves as a function of the observational noise level. We see that for $\Nobs=4$ VLKF always has a smaller RMS error than ETKF. \\
%For small noise levels the variance constraint is not used as frequently as for large noise levels due to $\Rw$

The results confirm again the results from our analysis of the toy model in Section~\ref{sec-toymodel}, that VLKF yields best performance for small observation intervals $\Delta \tobs$ and for large noise levels. For large observation intervals ETKF and VLKF perform equally well, since then the chaotic model dynamics will have lead the ensemble to have acquired the climatic variance during the time of propagation.\\

%

%\todo{Sensitivity analysis:\\
%I am confused: Fig3 of the old version where a non-constant mean was used and the new Fig3 (ie Fig7 from Lewis' draft), where $\amtar=\muclim={\rm const}.$, suggest that the results are highly sensitive to changes in mean and variance of target??? It is $20\%$ for $\Nobs=3$. There are 2 possibilities (1.) Fig3 and Fig 7 are not correct, but are actually compatible, or (2.) having some variance in the target means is a good thing to do. This needs to be checked. The sensitivity analysis so far assumes constant mean for all pseudo-observables.\\ (Also, Figs 3,4,5 are not really the umpf we were hoping for, I guess. If you can think of anything positive, let me know.)}

%\vspace{1cm}
In \citep{Ott04} it was observed that if not all variables $z_i$ are observed the Kalman filter diverges exhibiting blow-up. Similar behaviour was observed in \citep{HarlimMajda09}. In \citep{Ott04} the authors suggested that the sparsity of observations leads to an inhomogeneous background error, which causes an underestimation of the error covariance. We study here this catastrophic blow-up divergence (as opposed to filter divergence when the analysis diverges from the truth) and its dependence on the time between observations $\Delta \tobs$ and the proportion of the system observed $1 /\Nobs$. We note that blow-up divergence appears only in the case of sufficiently small observational noise and moderate values of $\Delta \tobs$. Once $\Delta \tobs$ is large enough (in fact, larger than the $e$-folding time corresponding to the most unstable Lyapunov exponent, in our case $2.1$ days) we notice that no catastrophic divergence occurs, independent of $\Nobs$. This probably occurs because for large observation intervals the ensemble acquires enough variance through the nonlinear propagation. We prescribe Gaussian observational noise of the order of $5$\% of the climatological standard deviation $\sigma_{clim}$, and set the observational error covariance matrix to $\Robs=(0.05\, \sigma_{clim})^2  \,\mathbf{I}$. The initial ensemble at $t=0$ is drawn again from an ensemble with variance $\sigmaclim^2$. 

To study the performance of VLKF when blow-up occurs in ETKF simulations we count the number $N_b$ of blow-ups that occur before a total of $100$ simulations have terminated without blow-up.  The proportions of blow-ups for the respective filters is then given by $N_b/(N_b+100)$. We tabulate this proportion in tables \ref{table:blowupVLKF} for the ETKF and VLKF respectively and the proportional improvement in table \ref{table:blowupimprovement}. The `x's' in the table represent cases where no successful simulations could be obtained due to blow-up.

%\todo{Can we get exact percentages for ETKF and VLKF for the "x"-cases? Is it 100\%? In Table 1 you mention $0.98$, which seems also severely. So what exactly means {\it severly} and the "x"'s??? {\it We note that in these cases the number of blowups was still far reduced using the VLKF}. Be specific! There should still be a difference between the filters? If you could not get any "successful $N+100$ cycle" for ETKF but for VLKF then this has to be documented with numbers for VLKF.}
%We note that in these cases the number of blowups was still far reduced using the VLKF.

Both filters suffer from severe filter instability for $\Nobs=6$, i.e. for very sparse observational networks, at small observation intervals $\Delta \tobs$. No blow-up occurs for either filter when every variable is observed. Note the reduction in occurrences of blow-ups for large observation intervals $\Delta \tobs$ as discussed above. We have checked that for all $\Nobs$ there is no blow-up for ETKF (and VLKF) for sufficiently large $\Delta \tobs$ (not shown); the larger $\Nobs$ the smaller the upper bound of $\Delta \tobs$ such that no blow-ups occur. Collapse is most prominent for ETKF (and for VLKF, but to a much lesser extent) for larger values of $\Nobs$ and at intermediate observation intervals which depend on $\Nobs$. Tables \ref{table:blowupVLKF} and \ref{table:blowupimprovement} clearly show that incorporating information about the pseudo-observables strongly increases the stability of the filter and suppresses blow-up. However, we note that despite the gain in stability VLKF has a skill less than the purely observational skill in the cases when blow-up occurs for ETKF, because the solutions become non-tracking. Further research is under way to improve on this in the VLKF framework. 

%We can measure the skill improvement of the VLKF for the case when the ETKF blows-up by looking at the RMS error in those cases. However, we find that in those cases the analysis for the pseudo-observables does not track the truth anymore. We can only define a skill improvement for the observed variables then...

%
\begin{table}
\begin{center}
\begin{tabular}{|r r||c|c|c|c|c|c|c|c|c|c|}
 \hline
 \multirow{6}{*}{$N_{obs}$ }\vline & 6 &   0.14 & x & x & 0.98 & 0.96 & 0.76 & 0.32 & 0.05 & 0.02 & 0.01\\
\vline& 5 &     0.02  &  0.40  &  0.67  &  0.73  &  0.84  &  0.89  &  0.94  &  0.82  &  0.49  &  0.19\\
\vline& 4 &     0    &     0.04  &  0.22   &   0.29  &  0.49  &  0.64  &  0.77  &  0.83  &  0.89  &  0.82\\
\vline& 3 &     0    &     0    &     0  &  0.03  &  0.04  &  0.11  &  0.15  &  0.44  &  0.58  &  0.67\\
\vline& 2 &     0     &    0     &    0    &     0    &     0  &  0.01    &     0  &  0.01    &     0.05   & 0.15\\
%\vline& 1 &     0    &     0    &     0    &     0    &     0    &     0    &     0    &     0   &      0   &      0\\
 \hline
 \hline
& & 0.025 & 0.05 & 0.075 & 0.1 & 0.125 & 0.15 & 0.175 & 0.2 & 0.225 & 0.25 \\
& & 3 h & 6 h & 9 h &  12 h & 15 h & 18 h & 21 h & 24 h & 27 h & 30 h \\
 \cline{3-12}
 & & \multicolumn{10}{|c|}{$\Delta \tobs$} \\
 \hline
% \end{tabular}
%\end{center}
%\caption{Proportion of catastrophically diverging simulations with ETKF. Observational noise with $\eta=0.05$ was used.}
%\label{table:blowupETKF}
%\end{table}
%
%\begin{table}[htdp]
%\begin{center}
%\begin{tabular}{|r r||c|c|c|c|c|c|c|c|c|c|}
 \hline
 \multirow{6}{*}{$N_{obs}$ }\vline & 6 &  0.01 & 0.42 & 0.11 & 0.01 & 0 & 0 & 0 & 0 & 0 & 0\\
\vline& 5 &     0  &  0.24  &  0.36  &  0.10  &  0.01    &     0    &     0    &     0    &     0     &    0\\
\vline& 4 &     0  &  0.03  &  0.22  &  0.12  &  0.06   &      0.02   &      0     &    0    &     0    &     0\\
\vline& 3 &     0    &     0    &     0  &  0.02  & 0  &  0.01  &  0.01  &  0.01  &  0  &  0\\
\vline& 2 &     0     &    0     &    0    &     0    &     0  &  0    &     0  &  0    &     0   & 0.01\\
%\vline& 1 &     0    &     0    &     0    &     0    &     0    &     0    &     0    &     0   &      0   &      0\\
 \hline
 \hline
& & 0.025 & 0.05 & 0.075 & 0.1 & 0.125 & 0.15 & 0.175 & 0.2 & 0.225 & 0.25 \\
& & 3 h & 6 h & 9 h &  12 h & 15 h & 18 h & 21 h & 24 h & 27 h & 30 h \\
 \cline{3-12}
 & & \multicolumn{10}{|c|}{$\Delta \tobs$} \\
 \hline
 \end{tabular}
\end{center}
\caption{Proportion of catastrophically diverging simulations with ETKF (upper table) and VLKF (lower table). Observational noise with $\Robs = (0.05\sigma_{clim})^2 \, \mathbf{I}$ was used.}
\label{table:blowupVLKF}
\end{table}
\begin{table}[htdp]
\begin{center}
\begin{tabular}{|r r||c|c|c|c|c|c|c|c|c|c|}
 \hline
 \multirow{6}{*}{$N_{obs}$ }\vline & 6 &  14  & x  &  x  & 98.00    &    $\infty$    &      $\infty$    &     $\infty$    &     $\infty$    &    $\infty$    &     $\infty$ \\
\vline& 5 &     $\infty$  &  1.67  &  1.86  &  7.30  &  84.00    &     $\infty$    &     $\infty$    &    $\infty$    &    $\infty$     &   $\infty$\\
\vline& 4 &     1  &  1.33  &  1.00  &  2.42  &  8.17   &      32.00   &      $\infty$     &   $\infty$    &     $\infty$    &    $\infty$\\
\vline& 3 &     1    &     1    &     1  &  1.5  & $\infty$  &  11.00  &  15.00  & 44.00  &  $\infty$  &  $\infty$\\
\vline& 2 &     1     &    1     &    1    &     1    &     1  & $\infty$    &     1  &  $\infty$    &    $\infty$   &  15.00\\
%\vline& 1 &     1    &     1    &     1    &     1    &     1    &     1    &    1    &     1   &      1   &      1\\
 \hline
 \hline
& & 0.025 & 0.05 & 0.075 & 0.1 & 0.125 & 0.15 & 0.175 & 0.2 & 0.225 & 0.25 \\
& & 3 h & 6 h & 9 h &  12 h & 15 h & 18 h & 21 h & 24 h & 27 h & 30 h \\
 \cline{3-12}
 & & \multicolumn{10}{|c|}{$\Delta \tobs$} \\
 \hline
 \end{tabular}
\end{center}
\caption{Proportional improvement of VLKF and ETKF as calculated as the ratio of the values from tables~\ref{table:blowupVLKF}.}
\label{table:blowupimprovement}
\end{table}
%

%\todo{Catastrophic divergence:\\
%We only need one information:\\
%Parameters ($\Nobs$, $\Delta \tobs$ and $\eta$) and skill improvement for case when\\
%ETKF has 100\% blow-ups\\
%VLKF has NO blow-ups, and less than 5\% (or so) do not track the solution.\\
%that way we get a reasonable skill improvement.\\
%This is extremely important!!!}

%Substantial improvements are seen using the VLKF when we incorporate information about the variance of the un-observed pseudo-observables. We set the variance of the pseudo-observables to be the climatic variance, and set $\atarget=\sigmaclim^2$. In table \ref{table:catdivVLKF} the proportion is shown, out of the same 100 realizations as those used to produced table \ref{table:catdivETKF}, which diverged catastrophically when filtering with the VLKF. Table \ref{table:catdivimprov} shows the proportional improvement of the VLKF over the ETKF in terms of reducing catastrophic divergence. In general, the VLKF shows larger improvements for less frequent obervations of the system, and for $N_{obs} = 4$. 

The fact that incorporating information about the variance of the un-observed variables improves the stability of the filter is in accordance with the interpretation of filter divergence of sparse observational networks provided in \citep{Ott04}. 
%In \citep{HarlimMajda09} it was suggested to replace the nonlinear propagation of the ensemble by a fitted linear stochastic process in order to avid filter divergence. For another recent method to control filter divergences, based on re-orthogonalization of the ensemble, see \citep{BGR09}.

%%%%%%%%%%%%%%%%%%%%%%%%%%%%%%%%%%%%%%%%%%

\section{Discussion}
\label{sec-disc}

We have developed a framework to include information about the variance of unobserved variables in a sparse observational network. The filter is designed to control overestimation of error covariances typical in sparse observation networks, and limits the posterior analysis covariance of the unresolved variables to stay below their climatic variance. We have done so in a variational setting and found a relationship between the error covariance of the variance constraint $\Rw$ and the assumed target variance of the unobserved pseudo-observables $\atarget$.\\

We illustrated the beneficial effects of the variance limiting filter in improving the analysis skill when compared to the standard ensemble square root Kalman filter. We expect the variance limiting constraint to improve data assimilation for ensemble Kalman filters when finite size effects of too small ensemble sizes overestimate the error covariances, in particular in sparse observational networks. In particular we found that the skill will improve for small observation intervals $\Delta \tobs$ and sufficiently large observational noise. We found substantial skill improvement for both observed and unobserved variables. These effects can be understood with a simple linear toy model which allows for an analytical treatment. We further established numerically that VLKF reduces the probability of catastrophic filter divergence and improves the stability of the filter when compared to the standard ensemble square root Kalman filter.\\

We remark that the idea of the variance limiting Kalman filter is not restricted to ensemble Kalman filters but can also be used to modify the extended Kalman filter. However, for the examples we used here the nonlinearities were too strong and the extended Kalman filter did not yield satisfactory results, even in the variance limiting formulation.\\

The effect of the variance limiting filter to control unrealistically large error covariances of the poorly resolved variables due to finite ensemble sizes may find useful applications. We mention here that the variance constraint is able to adaptively damp unrealistic excitation of ensemble spread in underesolved spatial regions due to inappropriate uniform inflation. This may be an alternative to the spatially adaptive schemes which were recently developed \citep{Anderson07,Li09}. In addition, it is known that localization of covariance matrices in EnKF leads to imbalance in the analyzed fields  (see, e.g., \cite{HoutekamerMitchell05,Kepert09}  for a recent study). Filter localization typically excites unwanted gravity waves which when uncontrolled can substantially degrade filter performance. One may construct balance constraints as pseudo-observations and thereby potentially reduce this undesired aspect of covariance localization. As more specific applications, we mention climate reanalysis and data assimilation for the mesosphere. It would be interesting to see how the proposed variance limiting filter can be used in climate reanalysis schemes to deal with the vertical sparcity of observational data and the less dense observation network on the southern hemisphere in the pre-radiosonde era (see \cite{Whitakeretal04}). One would need to establish though whether the historical observation intervals $\Delta \tobs$ are sufficiently small to allow for a skill improvement. Similarly, it may help to control the dynamically dominant gravity wave activity in the mesosphere as the upper lid is pushed further and further (see for example \cite{Polavarapu05}). However, a word of caution is required here. In some atmospheric data assimilation problems, it is not at all uncommon to have an ensemble prior variance for certain variables that is significantly larger than the climatological variance, when the atmosphere is locally far away from equilibrium. One relevant example would be in the vicinity of strong fronts over the southern ocean. 
%Even with radiance observations, it is still possible to have very large variance in temperature and wind at a gridpoint due to uncertainty in the exact position of a strong ``frontal" gradient. 
In such a case, it may not be appropriate to limit the variance to the climatological value.\\

In this work we have studied systems where for sufficiently large observation intervals $\Delta \tobs$ the variables acquire their true climatological mean and variance when the model is run. In particular we have not included model error. It would be interesting to see whether the variance limiting filter can help to control model error in the case that the free running model would produce unrealistically large forecast covariances. Usually numerical schemes do underestimate error covariances, but this is often caused by severe divergence damping \citep{durran} which is artificially introduced to the model to control unwanted gravity wave activity and to stabilize the numerical scheme. The stabiliziation may be achieved by a much smaller amount of divergence damping by implementing the variance limiting constraint in the data assimilation procedure. The VLKF would in this case act as an effective adaptive damping scheme, counteracting the model error.

%Shapiro, A., 1991: A note on the divergence damper as a means of noise control in a numerical hurricane prediction model. NOAA-NWS-NMC Office Note 384, 24 pp.

%%%%%%%%%%%%%%%%%%%%%%%%%%%%%%%%%%%%%%%%%%

\section*{Acknowledgements}
We thank Craig Bishop and Jeffrey Kepert for pointing us to the possible application of simulations involving the mesosphere. We thank the editor and three anonymous referees for valuable comments. GAG acknowledges support by the ARC.

%%%%%%%%%%%%%%%%%%%%%%%%%%%%%%%%%%%%%%%%%%

\addcontentsline{toc}{chapter}{Bibliography}
\bibliographystyle{natbib}
\bibliography{bibliography}

%%%%%%%%%%%%%%%%%%%%%%%%%%%%%%%%%%%%%%%%%%

\end{document}